\documentclass[sn-mathphys,Numbered]{sn-jnl}

\usepackage{graphicx}%
\usepackage{multirow}%
\usepackage{amsmath,amssymb,amsfonts}%
\usepackage{amsthm}%
\usepackage{mathrsfs}%
\usepackage[title]{appendix}%
\usepackage{xcolor}%
\usepackage{textcomp}%
\usepackage{manyfoot}%
\usepackage{booktabs}%
\usepackage{algorithm}%
\usepackage{algorithmicx}%
\usepackage{algpseudocode}%
\usepackage{listings}%
\usepackage{setspace}
\usepackage[T2A]{fontenc}

\theoremstyle{thmstyleone}%
\theoremstyle{thmstyletwo}%

\theoremstyle{thmstylethree}%

\begin{document}


\title[Article Title]{Deep Tissue Sensing of Chiral Molecules using Polarization Enhanced Photoacoustics}

\author[1]{\fnm{Swathi} \sur{Padmanabhan}}\email{swathip@iisc.ac.in}

\author*[1]{\fnm{Jaya} \sur{Prakash}}\email{jayap@iisc.ac.in}

\affil*[1]{\orgdiv{Department of Instrumentation and Applied Physics}, \orgname{Indian Institute of Science}, \orgaddress{\street{C. V. Raman Avenue}, \city{Bengaluru} - \postcode{560012}, \state{Karnataka}, \country{India}}}


\doublespacing

\abstract{Chiral molecule sensing is currently performed using chromatography, electrophoresis, enzymatic-assays, mass spectrometry, and chiroptical sensing techniques. Currently polarimetry is the only method having \textit{in-vivo} sensing capabilities, while other techniques analyze chiral molecules in body fluids. Polarimetry demonstrated \textit{in-vivo} performance upto depths of 1 mm while using UV-Visbile light, beyond which light scattering tends to be dominant. We hypothesize that photoacoustic sensing while operating at the Near-Infrared (NIR)-II window can allow for deep-tissue sensing, due to reduced scattering/autofluorescence effects. Herein, a Photoacoustic based Polarization Enhanced Optical Rotation Sensing (PAPEORS) system was developed for the first time to estimate optical rotation parameter using the recorded PA signals at larger depths. This optical rotation was then used to correlate with the chiral molecular concentration at depths of about 3.5 mm. Experimental evaluations were performed with aqueous glucose solution, naproxen drug, and serum-based glucose samples. PAPEORS has achieved a detection limit of 80 mg/dL using circular incidence polarization with serum-based glucose samples. Further \textit{ex-vivo} experiments were performed to demonstrate the ability of PAPEORS for deep-tissue sensing, which can be extended for \textit{in-vivo} chiral molecular sensing. PAPEORS uses single wavelength for sensing chiral biomolecules, hence can be miniaturized, allowing for many point-of-care applications.}




\maketitle

\doublespacing

\section{Introduction}\label{sec1}
Chirality is abundant in nature and plays a vital role in determining distinct properties of various organic and inorganic substances. 
Optically active molecules exhibit distinct properties compared to their enantiomeric counterparts\cite{marth2008unified,kelvin2021molecular}.
Many molecules present in the human body (proteins, amino acids, various sugars, and lactates) have a chiral centre, which enables them to rotate the plane of the incident polarized light. Inorganic molecules, mostly drugs like ketamine (anaesthesia drugs), steroids, and beta-blockers (to treat cardiovascular problems), were found to be chiral in nature\cite{h2011significance,kypr2009circular,morrow2017transmission}. The significance of utilising optical rotation has been demonstrated by various optical  techniques\cite{begin2023nonlinear,berova2007application} to either analyse the rotation of the molecules or for detection/estimation. 
Thalidomide is a drug used for treating morning sickness during pregnancy. Notably, administration of incorrect enantiomeric counterpart (S(+) thalidomide) resulted in severe birth defects, while R(-) thalidomide was found to be safe and used as a sedative\cite{therapontos2009thalidomide, BURLEY1962271}, this exemplified the importance of studying the effects of chiral molecules.
Consequently, pharmaceutical drugs continue to undergo scrutiny before being administered, drugs are analyzed in a neutralized (racemic) form or in a specific enantiomeric form to mitigate adverse effects. 

Conventional methods for sensing chirality include chromatography, electrophoresis, and mass spectrometry\cite{liu2023detection}. These are analytical detection methods and were calibrated for analysing the enantiomeric excess in a sample and for the estimation of chiral molecular concentrations. Note that all these methods rely on analyzing body fluids like blood, saliva, or urine, with less scope for {\it in-vivo} development. 
Chiroptical techniques considers the optical properties of chiral molecules like Circular Dichroism (CD), Raman Optical Activity (ROA), and Optical Rotary Dispersion (ORD) coupled with methods like spectroscopy\cite{habartova2018chiroptical,WANG2024115759}, and polarimetry\cite{he2021polarisation,ghosh2011tissue,lee2019digital}. 
A summary of the conventional and chiroptical methods highlighting various parameters and their feasibility in biosensing is shown in Table-S1. Currently, available chiroptical techniques pose many challenges for \textit{in-vivo} characterisation and quantification\cite{habartova2018chiroptical,liu2023detection,wolfender2015current}. 

Polarimetry\cite{stark2019broadband,li2021measuring,cameron1999use} was explored for biosensing to understand different structural and functional changes. Polarimetry has shown promise in the frontier of non-invasive imaging and treatment monitoring\cite{jacques2002imaging,qi2023surgical,lippok2017depolarization}.  Stokes Polarimetry, Mueller polarimetry, purity indices from the Mueller matrix, and optical rotation changes based on scattering were examined to determine the contrast difference between abnormal and normal tissues\cite{ghosh2011tissue,he2021polarisation}. Further polarimetry was used to detect changes in glucose concentration\cite{li2021measuring} and for studying structural changes in skin tissues\cite{jacques2002imaging}. The optical techniques based on polarization for chiral molecular sensing are limited by the penetration depth since most of the studies had used visible wavelength light operating at 589 nm and 632 nm\cite{co2004balanced,cote2005robust}, which is dominated by light scattering effects. CD-based techniques were employed to study protein structures and nucleic acids. CD-method was efficient in Ultraviolet-Visible (UV-VIS) range because the molecular electronic transitions were related to the absorption of circularly polarized light\cite{greenfield2006using,micsonai2015accurate}. Chiroptical spectroscopy methods were explored for wavelengths up to 1500 nm for photonic nanostructures\cite{kwon2023chiral}. Recently advanced methods for chiral sensing include Surface Enhanced Raman Scattering (SERS)\cite{arabi2022chiral} and Raman Optical activity\cite{palomo2022simultaneous,abdali2008surface}. Raman scattering-based methods were explored in the visible and infrared (up to 800 cm$^{-1}$) range for applications related to drug testing\cite{haesler2007absolute,parchavnsky2014inspecting}. Since, Raman methods  were known to suffer from low signal to noise ratio, SERS tagged the active molecules with metallic nanoparticles to increase the detected signal strength.  All chiroptical experimental configurations employed revealed information only from the superficial layers due to dominant scattering. Therefore, extracting information about the optical activity with respect to depth became quite challenging with polarization methods.

Photoacoustic (PA) sensing uses the principle of photoacoustic effect\cite{taruttis2015advances}  and has been used for multimodal cancer imaging and theranostic applications\cite{tripathi2023seed,liu2021croconaine}. 
Earlier works have demonstrated the advantages (like deeper penetration, less scattering and autofluorescence effects) of developing optical/photoacoustic systems while operating at the Near-Infrared-II (NIR-II; 1000-2000 nm) and Mid-IR window compared to visible/NIR region\cite{hong2014through,miao2018organic,hong2017near,shi2019high}.  NIR-II photoacoustic spectroscopy (PAS) was devised to estimate biomolecular concentrations using the acquired PA spectrum\cite{prakash2020short,ghazaryan2018extended}. Polarization properties like dichroism and Mueller matrix were investigated with Photoacoustic microscopy and Photoacoustic Computed Tomography(PACT)\cite{qu2018dichroism,zhou2019single} while operating in the Vis-NIR region. These studies enabled better contrast by using the concepts of linear dichroism and anisotropy. Another study\cite{zhang2023collagen}, integrated the concept of the Mueller matrix from the perspective of PA and demonstrated a better signal-to-noise ratio. However, the existing PA sensing/imaging methods do not consider optical rotation for concentration detection or sensing. Perceiving the challenges in chiroptical sensing and leveraging the advantages of photoacoustics, we have hypothesized that chirality-enhanced photoacoustic measurements could improve chiral biomolecular sensing, with potential \textit{in-vivo} application. 

In this work, we propose a method to estimate the optical rotation of chiral biomolecules in the context of photoacoustics, termed as - Photoacoustic based Polarization Enhanced Optical Rotation Sensing (PAPEORS). We analysed the PA time series data obtained in a transmission mode configuration for sensing concentrations of optically active substances like glucose and naproxen (NSAID-Non-steroidal anti-inflammatory drugs used for relieving pain and muscle aches). 
This study investigated the potential of PAPEORS for the detection of chiral molecules in deep tissue and to address challenges faced during \textit{in-vivo} chiral molecular sensing. To the best of our knowledge, the investigation of optical rotation changes by detecting the PA signals was not explored. PAPEORS was used to: 
\textbf{(i)} estimate the optical rotation from PA time-series measurement using a trans-illumination configuration, \textbf{(ii)} validate the proposed system for accurately sensing chiral biomolecular concentration using optical rotation at larger depths i.e. about 3.5 mm, \textbf{(iii)} evaluate the optical rotation changes with respect to different polarized incidences: Vertical(V), 45$^\circ$ Linear(P), and Circular(R) polarization, and  \textbf{(iv)} conceptualize a non-invasive polarized photoacoustic sensing method. This work evaluated the variation in the estimated optical rotation (from PAPEORS) as a function of chiral molecular concentration in aqueous-based glucose samples, serum (Bovine Serum Albumin-BSA) based glucose samples, and Naproxen. Lastly, we discussed the prospects of PAPEORS by highlighting a single wavelength usage which can lead to easy miniaturization of the system.

\begin{figure}
\centering
\includegraphics[width=1\linewidth]{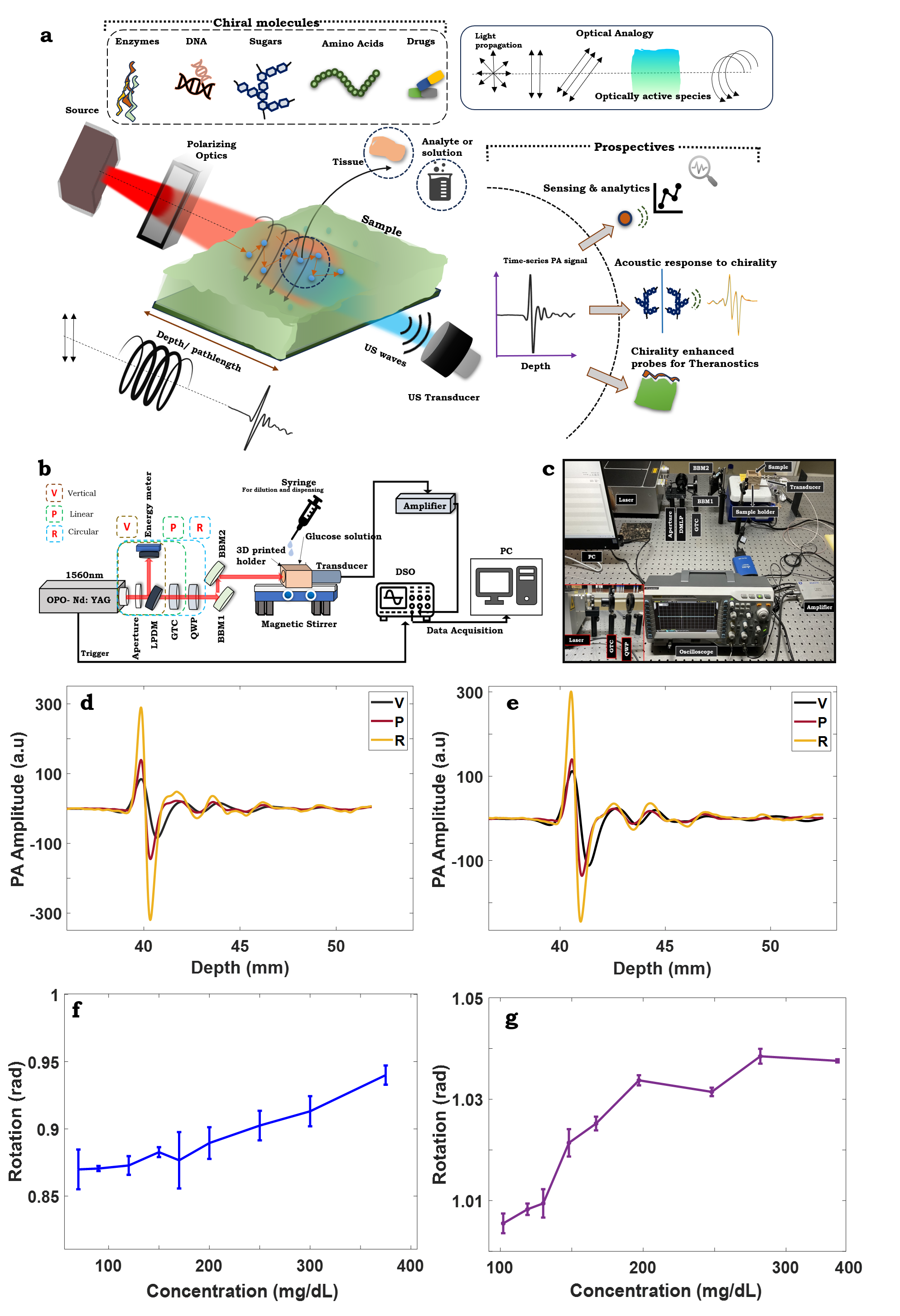}
\rule{0pt}{6ex}
\caption{\textbf{Principle of PAPEORS: (a)} Schematic describing the concept and potential applications of PAPEORS. \textbf{(b)} Schematic of the experimental setup used for optical rotation measurements. The components used for the optical alignment and polarization generation are an aperture of diameter 6 mm to reduce beam diameter, \textbf{DMLP}: Long Pass Dichroic Mirror, \textbf{ GTC}: Glan Taylor Crystal Polarizer, \textbf{QWP}: Quarter Wave Plate, and two \textbf{BBM1 \& BBM2}: Broad Band Mirrors. Energy meter was aligned with \textbf{DMLP} for real-time energy measurements. \textbf{(c)} shows the photograph of the experimental setup used for acquiring the photoacoustic measurements. \textbf{ (d)} and\textbf{ (e)} show the acquired PA signals for the aqueous and BSA glucose solutions, respectively, for V, P, and R incidences for 350 mg/dL concentration. The R incidence shows the best SNR for both samples compared to V and P incidence.  \textbf{ (f)} and  \textbf{ (g)} shows the optical rotation analysed at a depth of 2 mm from the PA experiments for V incidence. \textbf{ (f)} shows the optical rotation as a function of concentration for aqueous glucose samples and \textbf{ (g)} for the serum based samples. }
\label{<figure-label>}
\end{figure}

\section{PAPEORS Principle}
The optical activity of the chiral molecule is measured in terms of the rotation of the molecule. The optical rotation of the molecule depends on the incident state of polarization, concentration (\textit{c}), path length (\textit{l}), and the wavelength ($\lambda$) of light. The relation between concentration and optical rotation ($\theta$) is given as\cite{lide2004crc},
\begin{equation}
    c = \dfrac{\theta}{[\theta]_\lambda^T * l}
\end{equation}
where $[\theta]_\lambda^T$ is the specific rotation of the chiral molecule and is dependent on wavelength and temperature. The pressure generated by a chromophore as a result of the absorption at a certain wavelength using the PA effect is given as\cite{prakash2020short},
\begin{equation}
        P = \tau \mu_a \phi
\end{equation}
The generated PA signal is proportional to the Gr\"uneisen parameter ($\tau$), absorption coefficient  ($\mu_a$), and light fluence ($\phi$). We propose a method to combine and study the optical rotation of glucose/naproxen by inferring the fluence from the recorded PA time-series signals. For combining the optical and photoacoustic phenomena, we have considered the Malus' Law, given as,
    \begin{equation}
         I = I_0cos^2 \theta 
    \end{equation}
Generally, \textit{I} is considered as the intensity transmitted and $I_0$ is the incident light intensity. This work has correlated the light fluence  at different depths to the time-series PA signal, such that it is analogous to the Malus' law representation, as follows,
    \begin{equation}
        P = P_0cos^2 \theta 
    \end{equation}  
where $P$ and $P_0$ denote the amplitudes of PA signal and are chosen based on the depth or the path length of the light travelled. Finally, the rotation ($\theta$) experienced by the chiral molecules can be computed from the above equation once P and $P_0$ are known using,
    \begin{equation}
         \theta = cos^{-1}\sqrt{\dfrac{P}{P_0}} 
    \end{equation}
The recorded PA signal can be directly correlated to the light fluence (with an assumption of having a constant Gr\"uneisen coefficient) due to lesser scattering effects at the NIR-II window. 

 $P_0$ was chosen close to the illumination point on the holder (approx 0.018 mm), and $P$ was chosen at a particular depth depending on the path length of interest (1.5 mm, 2 mm, or 3 mm). The detailed steps involved in extracting the points from deconvolution are described in Fig. S1. 
Based on the optical theory, rotation is observed to increase linearly with the concentration\cite{ghosh2011tissue,barron2009molecular}. However, we observed a slightly nonlinear behaviour for the rotation at lower concentrations (i.e., from 70 to 160 mg/dL) and good linearity for higher concentrations (Refer to Figs 1(f)-(g)). Hence, a quadratic regression model was used for concentration estimation. The details of the model used for the concentration prediction ($C_p$) obtained from the observed rotation ($\theta$) is explained in the Section 5.2.

\section{Results}\label{sec2}
\textbf{Glucose rotation estimation using PAPEORS} \\
Photoacoustic signals with three different polarization incidences for optical rotation estimation were obtained using the setup shown in Fig. 1(c). Details of the experimental system is elaborated in Section 5.1.
The raw PA signals (acquired at 1560 nm) for aqueous glucose sample and BSA-based sample having a concentration of 350 mg/dL for different polarized incidences are indicated in Figs 1(d) and 1(e) respectively.
After energy normalization, the signal amplitudes for circular and linear polarization states had a higher signal-to-noise ratio than those for vertical incidence. The recorded PA signals were affected by the light attenuation and the transducer response. 
The deconvolved time-series PA signal (as explained in Fig. S1 \& Supplementary text II) was used to estimate the optical rotation.  

\textbf{PA spectrum for V, P, and R polarized incidence} \\
Three concentrations within the physiological range (100, 200, and 300 mg/dL) and one highly saturated concentration (2000 mg/dL) were chosen for analyzing the PA spectrum with different polarization incidences. The spectra shown in Figs 2(a)-(c) were for serum (BSA)-based samples for V, P, and R incidences. PA measurements were also performed for the aqueous glucose solutions (Fig. S2). The baseline spectrum of water and BSA were recorded separately to enable baseline correction. The glucose peak absorbance was observed from 1500 nm for vertical, linear and circular incidences. 
Figs 2(a)-(c) clearly illustrate that the PA signal does not show a linear correspondence with an increase in glucose concentration, i.e. the recorded PA signal for the highly saturated concentration (2000 mg/dL) was found to have lesser intensity than 300 mg/dL in all the instances and with multiple repetitions. 
PA amplitude as a function of concentration at 1.7 mm depth (from the point of illumination) is represented in Figs 2(d)-(f). Due to light attenuation effects, the recorded PA signals do not linearly increase with concentration; note that such non-linear behaviour was found at depths larger than 1 mm. Fig. 2 clearly demonstrates that the linearity fails from spectroscopic techniques after 1.7 mm depth for the serum-based samples. Figs 2(d)-(f) show the non-linearity for depths greater than 1.7 mm with linear and circular incidence. Similar variations for the aqueous glucose samples are reported in Figs S2(d)-(f). Hence, we concluded that PA spectroscopy might fail at larger depths, making it difficult to extract the chiral molecular concentration reliably from the recorded PA data. 


\begin{figure}
\centering
\includegraphics[width=1\linewidth]{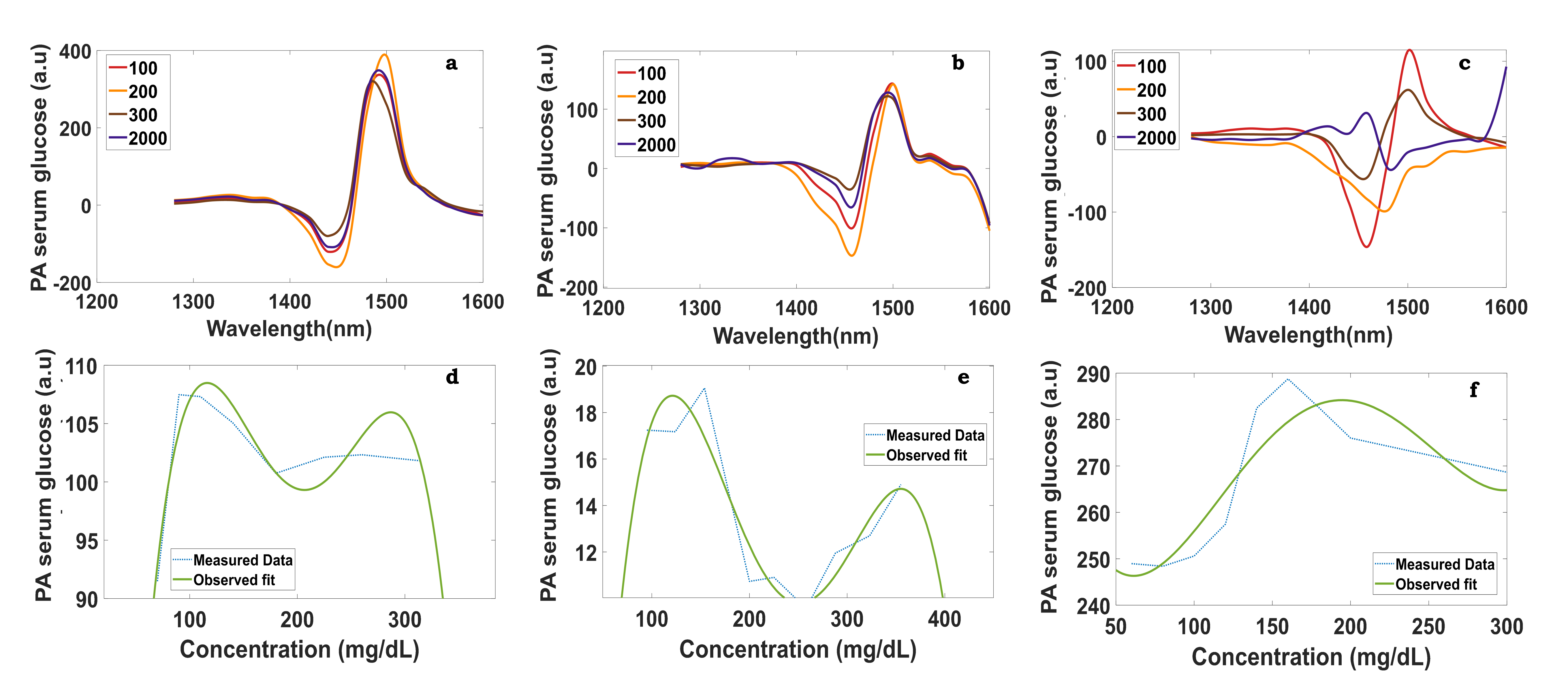}
\rule{0pt}{6ex}
\caption{\textbf{Photoacoustic spectrum and non-linear effects with serum based samples:} PA Spectra for the \textbf{(a)} Vertical (V) incidence, \textbf{(b)} Linear (P) incidence, and \textbf{(c)}  Circular (R) incidence.  \textbf{(d), (e), } and \textbf{(f)} shows the non-linear variation of the PA amplitude as a function of concentration at a depth of 1.7 mm for V, P, and R incidences, respectively.}
\label{<figure-label>}
\end{figure}

\textbf{Optical rotation of D-Glucose from Photoacoustic measurements} \\
D(+) glucose samples (prepared in aqueous and BSA solution) are dextrorotatory in nature. 
The recorded PA signal was initially pre-processed and then deconvolved with the transducer response to give a final output signal, which clearly shows the light attenuation signal (Fig. S1(d)). The PA amplitudes were then chosen as described in Section 2 and Fig. S1. These amplitudes, i.e., $P$ and $P_0$, were then used to compute the optical rotation.
Optical rotation as a function of concentration at a depth of 2 mm for the V incidence is illustrated in Figs 1(f) and 1(g). Fig. S3 shows the calculated rotation varying as a function of concentration for P and R incidences for aqueous solution (Fig. S3(a)-(b)) and serum-based samples (Fig. S3(c)-(d)). The changes in rotation observed are in the order of millidegrees to degrees, which aligns well with polarimetric methods\cite{cote2005robust}. 

A linear relationship between rotation and concentration was observed, which corresponds well with optics literature\cite{barron2009molecular}. 
Factors affecting rotation include: (a) water absorption interfering with PA signal, (b) pathlength choice, (c) laser energy fluctuations, (d) solvent-based Brownian motion, and (e) experimental geometry. Owing to these factors, data was collected over multiple days to ensure repeatability (Fig. S6(a) and (b)).
Further, Polarized Monte Carlo\cite{ramella2005three} simulations were performed to validate the obtained experimental optical rotation estimates from different polarized incidences. We observed a good correlation between the simulated fluence profiles and experimental data obtained at a depths of 1.7 mm with different polarization incidences (Fig. S4).

\textbf{Concentration estimation from the rotation of D-Glucose} \\
A quadratic regression model was used as indicated in Eq. 6.  Approximately 3 outliers per dataset were eliminated while building the model during the analysis. Data acquired from the same day was used to build the model and perform the estimation. Fig. 3 shows the Clarke's Error Grid Analysis (CEGA) for the aqueous and serum-based glucose samples with V, P, and R  incidences. 
The concentration estimation (with serum-based samples) from the vertical incidence had about 87\% points in the Zone-A while linear incidence had about 79\% points. The circular incidence had 82\% of the total points lying in Zone-A of CEG. The important parameters derived are summarised in Table-1. Few estimated points for the aqueous samples case fell in Zone-B, because the estimated rotation had high variance as a result of dominant water absorption. The accuracy of the model could be further improved using advanced machine learning models.

The limit of detection (LOD) for the predicted glucose levels was determined for each polarized incidence based on points falling in Zone-A of CEG. Notably, the detection limit for V incidence data is the worst at 120/107 mg/dL compared to P and R incidences for aqueous/BSA samples. 
The circular incidence data yielded the lowest predicted limits for aqueous and BSA samples at 80 mg/dL and 81 mg/dL. \\

\begin{figure}
\centering
\includegraphics[width=1\linewidth]{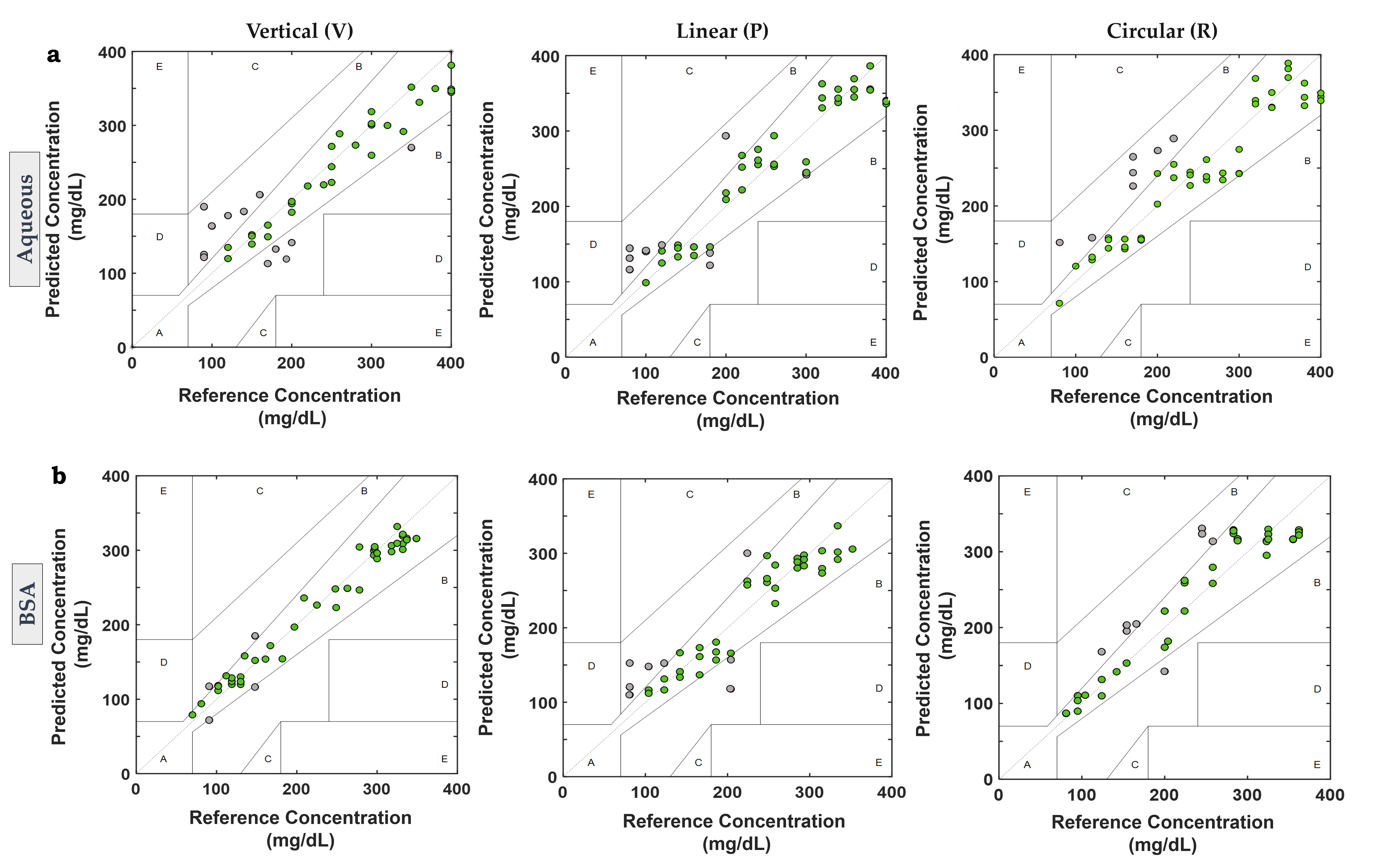}
\rule{0pt}{6ex}
\caption{\textbf{Clarke's Error Grid Analysis for Glucose estimation using PAPEORS:} \textbf{(a)} Top row shows the CEGA for the aqueous glucose samples for V, P, and R polarized incidences respectively, and \textbf{(b)} Bottom row shows the CEGA for  serum-based glucose samples for the V, P, and R polarized incidences respectively. The BSA glucose samples for the vertical incidences shows the highest percentage of predicted points in Zone-A. Overall, serum samples show a higher percentage of predicted concentrations in Zone-A than the aqueous samples for all three polarized incidences.}
\label{<figure-label>}
\end{figure}

\textbf{\textit{Ex-vivo} PAPEORS experiments for deep-tissue glucose detection} \\
\textit{Ex-vivo} PA experiments were performed using thin slices of chicken breast. The details of the sample preparation are elaborated in Section 5.4. The configuration used for data acquisition can be seen in Figs 4(a) and 4(b). The tissue slices were placed perpendicular to the axis of illumination and were held by a fixture at the bottom of the holder. The space between the two slices was filled with a serum-based glucose solution having different concentrations. 
This kind of configuration was attempted to mimic an \textit{in-vivo} scenario from where we can detect the blood glucose levels inside thick biological tissue samples. The thickness of the slice placed close to the illumination source (Fig. 4(a)) was varied, i.e. 2 mm and 3.5 mm, to assess the system's penetration and its ability to generate acoustic signals. PA signals for 10 glucose concentrations from 90 to 400 mg/dL were recorded at 2 mm and 3.5 mm depths. Firstly, the rotation was calculated, and then the concentrations were predicted using the quadratic fitting model (Eq. 6) for V, P, and R incidence configurations. The predicted glucose concentrations from the \textit{ex-vivo} experiments can be seen in Figs 4(c) and 4(d) for the V incidence at 2 mm and 3.5 mm, respectively. 89.5\% of predicted concentrations fell in Zone-A of the CEG for the vertical incidence at both the depths. CEGA for the P and R incidence configurations is shown in Fig. S7. Table-1 summarizes the parameters observed from the CEGA for \textit{ex-vivo} experiments with all three polarized incidences using PAPEORS. 

\begin{figure}
\centering
\includegraphics[width=1\linewidth]{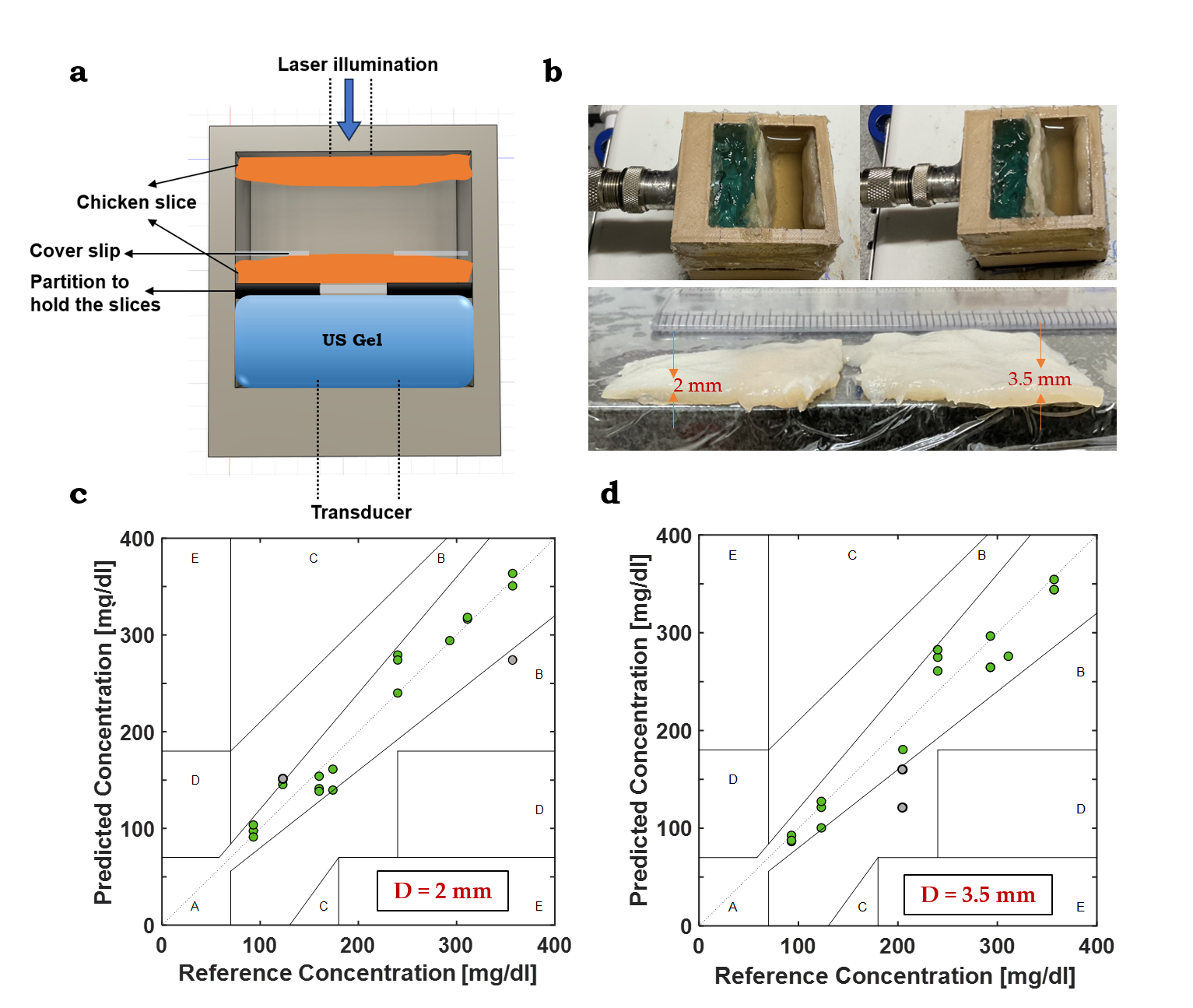}
\rule{0pt}{6ex}
\caption{\textbf{Ex-vivo studies for serum based glucose concentration estimation: (a)} shows the schematic of the holder and the positioning of two chicken breast slices during PA measurements at two different depths: 2 mm and 3.5 mm, \textbf{(b)} shows the dissected slices used for the experiments and placements of the same during measurements, \textbf{(c)} and \textbf{(d)} show the CEG for the vertically polarized incidence for 2 mm and 3.5 mm respectively. }
\label{<figure-label>}
\end{figure}

\begin{table}[]
    \centering
\begin{tabular}{|c|c|c|c|c|c|c|}
\hline Incident Polarization & \multicolumn{2}{|c|}{ \textbf{VERTICAL} } & \multicolumn{2}{|c|}{ \textbf{LINEAR} } & \multicolumn{2}{|c|}{ \textbf{CIRCULAR} } \\
\hline \hline \textbf{Solution } & Aqueous & $B S A$ & Aqueous & $B S A$ & Aqueous & $B S A$ \\
\hline Zone A \% & $62.5 \%$ & $\mathbf{\textcolor{blue}{8 7 . 5 \%}}$ & $\textcolor{blue}{\textbf{87 \%}}$ & $79.3 \%$ & $82 \%$ & $84.4 \%$ \\
 LOD & $120 \mathrm{mg} / \mathrm{dL}$ & $107 \mathrm{mg} / \mathrm{dL}$ & $98 \mathrm{mg} / \mathrm{dL}$ & $104 \mathrm{mg} / \mathrm{dL}$ & \textbf{\textcolor{blue}{80 mg/dL}} & $\textcolor{blue}{\textbf{81 mg/dL}} $ \\
\hline\hline  \textit{\textbf{Ex}}-\textit{\textbf{vivo}}  & \multicolumn{2}{|c|}{$B S A$} & \multicolumn{2}{|c|}{$B S A$} & \multicolumn{2}{|c|}{$B S A$} \\
\hline 2mm Zone A \% & \multicolumn{2}{|c|}{$\textcolor{blue}{\textbf{89.5 \%}}$} & \multicolumn{2}{|c|}{$85 \%$} & \multicolumn{2}{|c|}{$85 \%$} \\
  LOD & \multicolumn{2}{|c|}{$91 \mathrm{mg} / \mathrm{dL}$} & \multicolumn{2}{|c|}{$101 \mathrm{mg} / \mathrm{dL}$} & \multicolumn{2}{|c|}{$\textcolor{blue}{\textbf{88 mg/dL}}$} \\
\hline 3.5mm Zone A \% & \multicolumn{2}{|c|}{$\textcolor{blue}{\textbf{89.47 \%}}$} & \multicolumn{2}{|c|}{$75 \%$} & \multicolumn{2}{|c|}{$70.5 \%$} \\
  LOD & \multicolumn{2}{|c|}{$\textcolor{blue}{\textbf{93 mg/dL}}$} & \multicolumn{2}{|c|}{$100 \mathrm{mg} / \mathrm{dL}$} & \multicolumn{2}{|c|}{$94 \mathrm{mg} / \mathrm{dL}$} \\
\hline
\end{tabular}
 \caption{\textbf{Estimated Parameters for Ex-vivo experiments and CEGA of glucose solutions:} The percentage of predicted concentrations falling in the A-Zone of the CEGA and the estimated Limit of Detection, \textbf{LOD} is shown for the aqueous glucose, serum glucose samples and\textit{ ex-vivo} chicken tissues. The parameters are summarised for V, P, and R incident configurations. The best LOD was observed for circular incidence for the solution samples. The best predicted concentrations are seen for the Vertical incidence data and the lowest LOD for circular incidence at a depth of 2 mm for the \textit{ex-vivo} experiments. }
    \label{tab:my_label}
\end{table}

\textbf{Optical rotation of Naproxen using PAPEORS} \\
Naproxen is an NSAID drug used to treat mild to severe pain and inflammation. This molecule is known to exhibit D+ rotation\cite{o2013merck}. The spectrum was characterised for Naproxen at three different incidences to identify the absorption peak. Post the blank correction with ethanol, an absorption peak was observed at 1500 nm. The recorded spectra for Naproxen with V, P, and R incidences are given in Fig. S5. Rotation experiments were performed by diluting a high concentration Naproxen, i.e. 2.8 mg/mL with 70\% ethanol. 
Fig. 5 shows the variations in optical rotation observed with changes in Naproxen concentration using PAPEORS. Figs 5(a)-(c) show the rotation trends observed with different polarized light incidences. The rotation for Naproxen was calculated at a depth of 0.5 mm, beyond which fluctuations started to amplify. 
Since the optical rotation was varying linearly with concentration, a single linear equation was used to build the fitting model, unlike the glucose case. The reference concentration was considered based on the prepared weight-to-volume ratio. The predicted concentrations (using the linear fitting model) were plotted against the reference concentrations, and the goodness of fit was evaluated, as shown in Figs 5 (d)-(f). The $R^2$ values are also shown as the inset of Fig. 5. 

\begin{figure}
\centering
\includegraphics[width=1\linewidth]{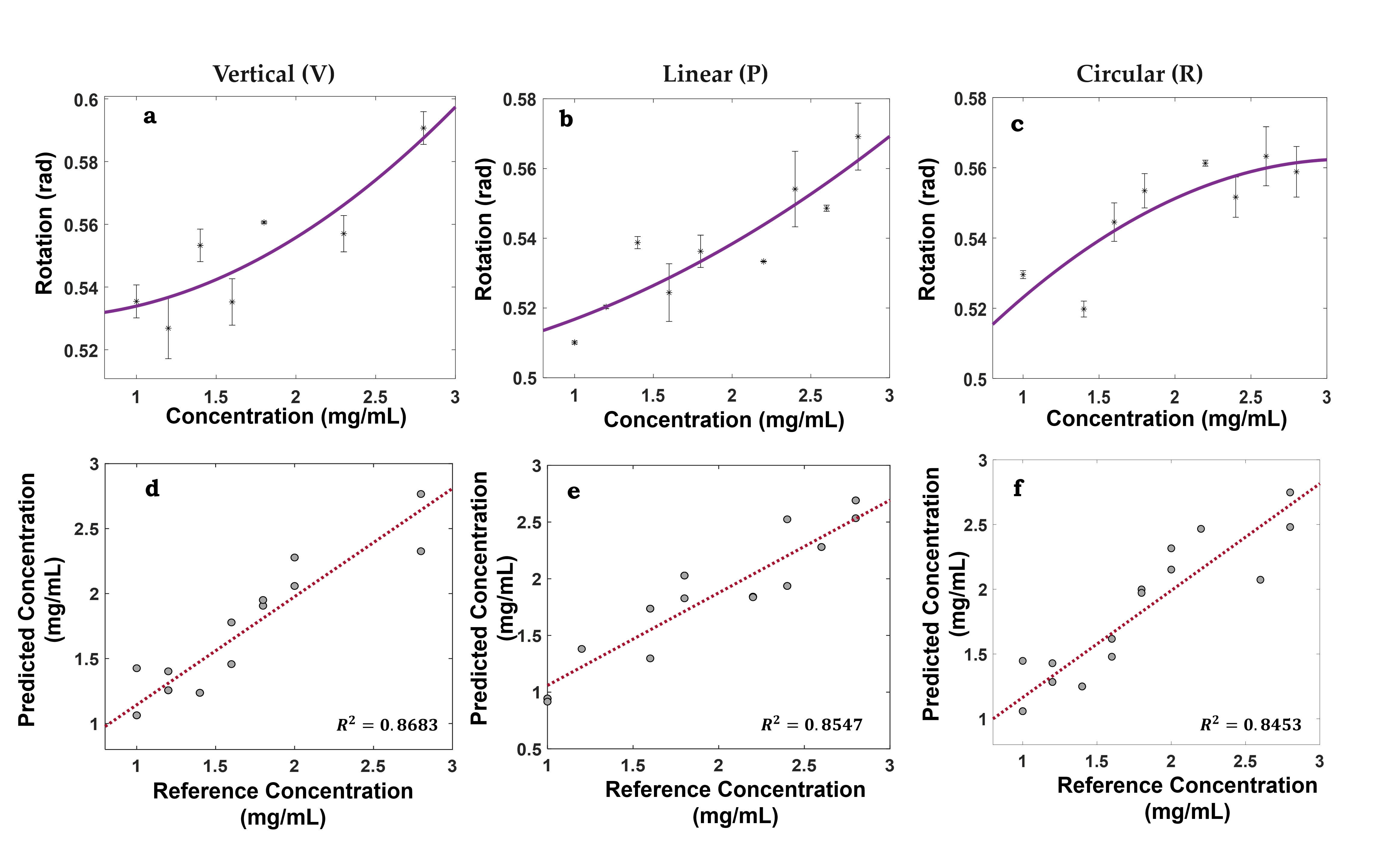}
\rule{0pt}{6ex}
\caption{\textbf{Optical Rotation and Concentration Predication of Naproxen using PAPEORS:} \textbf{(a), (b)}, and \textbf{(c)} showcases the variations in V, P, and R incidences for the Naproxen solution having the concentration range  of 0.8 to 2.8 mg/mL. \textbf{(d), (e)}, and \textbf{(f)} show the predicted concentration using linear regression based on the variations in the rotation. The $R^2$ values are indicated as inset in (d)-(f).  }
\label{<figure-label>}
\end{figure}

\section{Discussion}
PAPEORS has the ability to measure the optical rotation at unprecedented depths in biological tissue/turbid media. Our work revealed that there are variations in the observed PA signals when the incident state of light polarization was changed. 
CEGA for glucose sensing suggests the potential application of PAPEORS for continuous blood glucose monitoring, which is critical for diabetes management\cite{li2023review}. Photoacoustic imaging was indeed used earlier for direct glucose sensing\cite{ghazaryan2018extended}, and assessing diabetes-related risk factors by studying vascular structures\cite{karlas2023dermal}. PAPEORS can be customized to provide direct \textit{in-vivo} glucose information making it a strong candidate to solve the holy grail in diabetes management.
PAPEORS can be developed as a straightforward, non-invasive system, which will not require complex sample preparation steps like the conventional chiral sensing methods\cite{liu2023detection}. Chiral molecular sensing using PAPEORS was found to be more accurate than photoacoustic spectroscopy (due to non-linearities) at larger depths. 
The time-series PA measurements allows us to precisely sense from a specific region/depth of the tissue. PAPEROS sensed glucose at a depth of 2 mm (with chicken breast slice) having LOD as 88.5 mg/dL using circularly polarized light incidence. Table-1 shows the estimated parameters for \textit{ex-vivo} experiments, which indicates PAPEORS' effectiveness even with complex biological structures. We were able to estimate concentration levels from 80 mg/dL with our proposed quadratic regression model without any complex algorithms. Further advanced machine learning methods can be integrated as part of PAPEORS to improve sensitivity and prediction accuracy.

We have overcome the scattering effects by using NIR-II wavelengths i.e. 1560 nm and 1500 nm that allows deeper penetration. Currently, PAPEORS employs a single wavelength for concentration estimation, which is highly advantageous in miniaturising the system for a focused sensing application. The biological fluids contain more than one functional molecule of interest. Therefore, signal interference from these molecules like water, lipids and proteins might affect the measurements. Hence, it is recommended to perform initial spectroscopy which can reveal the wavelength having higher absorption corresponding to specific moiety of interest. We have indeed demonstrated realistic \textit{ex-vivo} experiments with serum-based samples, indicating PAPEORS' potential to sense chiral molecules with limited interference. Experiments were repeated multiple times to ensure repeatability. 

PAPEORS analysed the optical rotation from PA signals and demonstrated a good sensing method, however it was not devoid of limitations. Firstly, (i) Serum-based glucose sample was investigated as they closely mimic the blood plasma. Further \textit{ex-vivo} configuration was attempted to replicate a controlled environment compared to \textit{in-vivo} setting (as glucose concentration is known to vary drastically \textit{in-vivo}). These experiments gave foundational insights into the optical rotation estimates from PAPEORS. However \textit{in-vivo} studies with PAPEORS should be performed to evaluate its reliability. Secondly, (ii) Multiple chiral molecules are present in the human body and are significant biomarkers. Characterising them and sensing them simultaneously was not explored using PAPEORS. PAPEORS was indeed used to demonstrate sensing of multiple chiral molecules by using different wavelengths (corresponding to the absorption peak of each moiety). Hence, it is a straightforward extension of PAPEORS to study mixtures of chiral molecules.
PAPEORS sensing can be extended to investigate and quantify the specific rotation of chiral particles, circular dichroism, and birefringence at a deep tissue level and explore the impact of incident polarized light rotation on the generated acoustic signals. Currently, PAPEORS was evaluated at NIR-II wavelengths, however similar concepts can be explored at other wavelengths like mid-IR, terahertz, etc.
PAPEORS has potential for compact wearable integration\cite{westerveld2021sensitive} and coupling with endoscopic probes for high resolution PA imaging enabling gastrointestinal cancer/lesion detection\cite{liang2022optical}. Since, PAPEORS offer deeper penetration of light and good contrast, it holds promise in the context of multi-modal imaging and combined therapies. 

\section{Methods}\label{sec11}
\subsection{Photoacoustic experimental setup} 
The optical rotation of the glucose samples were studied using the setup shown in Fig. 1(b). The glucose solution placed in a custom-designed 3D printed holder was illuminated using a nanosecond pulsed Nd:YAG laser (Innolas SpitLight OPO, Wavelength 1100-1700 nm, 30 Hz, 7 ns). The photoacoustic signals were recorded using a single element ultrasound transducer having a central frequency of 7.5 MHz (Focused, Olympus V320-SU-F1.2IN-PTF, 60 $\%$ bandwidth) fixed in the 3D printed holder containing the sample. The focus of the transducer used is 1.2 inches. The schematic in Fig. 1(b) shows the configuration used for acquiring the optical rotation measurements. The rotation experiments were performed in a trans-illumination configuration. An aperture of 6 mm diameter was positioned at the laser output to reduce the beam diameter. The beam diameter was reduced to be within the damage threshold values of the polarization optics used in the experiments for alignment. A long pass dichroic mirror from ThorLabs (DMLP650) was used in the beam splitting configuration. An external energy meter (Newport Pyroelectric energy sensor) was used to record the energy falling on the sample for each incident state of polarization. The energy meter was kept in the path of the reflected ray from the DMLP and the transmitted ray through DMLP was allowed to pass through the rest of the optics and then to the sample. The acquired time-series PA signals were normalised with the energy measured in real-time using an external energy meter.  A total of 60 (30 on the same day and 30 on the second day) readings were acquired on the same day and multiple days to ensure the reliability of the setup designed for V, P, and R incidences (Fig S6(a) and (b)).

The laser beam was further aligned using two Broad Band Mirrors -\textbf{BBM} (Thorlabs, BB1- E04) inclined at 45$^\circ$ to each other to adjust the working height for the data acquisition configuration.  Different states of polarization was considered as input, initially, a vertically polarized light was directly used as an illumination source for PA measurement acquisition. In the second configuration, the vertically polarized light was used to generate a linearly polarized light, this was achieved by placing a Glan Taylor Crystal (\textbf{GTC}) polarizer (Thorlabs GT-10C), for introducing a phase shift to the originally vertically polarized light of the laser to become linearly polarized illumination source.  Final configuration involved generation of circular polarized light (as illumination source) by placing a quarter wave plate (\textbf{QWP}), Thorlabs AQWP05M-1600 in front of the\textbf{ GTC}. The transducer was positioned at a depth of 4.1 cm from the point of illumination fixed with silica gel to prevent leakage of the glucose solution. The illumination side of the holder was covered with a cover slip and silica gel. The time-series PA signals were collected using a preamplifier (Olympus, PR-06-04) having a 40 dB gain through an oscilloscope (RIGOL MSO2102A). Post-recording, the signals were processed by filtering and a series of operations, which are described in the next section. The energy per pulse at 1560 nm from the laser was 2.5 mJ, 1.18 mJ and 1 mJ for V, P and R, respectively. The peak energy densities from the laser was calculated to be 0.04J/$cm^2$, 0.02 J/$cm^2$, and 0.015 J/$cm^2$ for V, P, and R incidence respectively, which is within the ANSI limits and did not cause any damage to the tissue used for the \textit{ex-vivo} studies. 
\\

\subsection{Post Processing and Concentration Estimation} 
The PA signals recorded during experiments were processed over a series of operations for both spectra and optical rotation estimation. For the spectral data case, the PA signals were filtered using a Chebyshev filter with bandpass limits of 1 MHz - 9 MHz. Peak-to-peak amplitude of the PA signal was considered across the wavelength range of 1280 nm to 1600 nm. Baseline spectra of water and BSA were subtracted for each sample. The energy recorded at different wavelength during the experiments was used to normalize the PA spectrum data. \\

The time series data collected for the rotation measurements were also filtered using the same Chebyshev filtering parameters. The speed of sound (SoS) is crucial in establishing the correlation between time and depth. A brief assessment of SoS was conducted to determine and correlate the actual physical depth of the custom holder. The adjusted SoS values were 1420 m/s for the solution-based glucose samples and 1490 m/s for the \textit{ex-vivo} chicken tissue experiments. The discrepancy is due to variations in medium properties. The filtered signals were observed to be convolved output of the PA signal with the transducer response. A deconvolution method was opted to correct for the transducer response from the recorded experimental PA measurements. The steps involved in the process of deconvolution are shown in Fig. S1. The quadratic regression model for predicted concentrations ($ C_{p}(\theta)$) based on the rotation vs concentration profile was:
\begin{equation}
  C_{p}(\theta) =
    \begin{cases}
      a_1\theta^2 + b_1\theta + k_1, & { \theta \leq \theta_T } \\
      a_2\theta^2 + b_2\theta + k_2, & \theta > \theta_T \\
    \end{cases}       
\end{equation}
where $\theta_T$ is the threshold value for rotation. $a_1$, $b_1$, $k_1$ and $a_2$, $b_2$ and $k_2$ are the coefficients of the quadratic equations being utilised for the concentration prediction. The values were decided based on the variations observed in the rotation detected from the photoacoustic measurements for a given path length and the range of concentrations. 
The value of $\theta_T$ was chosen close to 200 mg/dL with rotation magnitudes of 0.88, 0.98, and 0.822 rad for serum samples, and 0.88, 1.025 and 0.97 rad for aqueous glucose samples.  The predicted concentrations ($C_p$) were obtained from the observed rotation ($\theta$). All the signals acquired were normalised with the energy values measured using the energy meter. All the scripts for analysis and post-processing were written using MATLAB R2020b. Note that a linear model was used for estimating the concentration in the case of Naproxen.
\\

\subsection{Glucose Sample preparation} 
Weight-to-volume metric calculations were used to estimate the weight of D+ glucose (SIGMA-CAS 50-99-7) required to be added. The aqueous glucose solutions were prepared by dissolving the weighted glucose in deionised water using a magnetic stirrer plate. Serum-like samples were prepared by dissolving BSA (Himedia, fraction V BSA) in deionised water and then adding the weighed glucose corresponding to each concentration to mimic the nature of blood plasma. The concentration of BSA prepared was 4 g/dL, which falls under normal human serum albumin levels (Normal range: 3.4 g/dL - 5.4 g/dL). The glucose concentration range was chosen to be from 50 mg/dL to 400 mg/dL, which lies within the physiological range and one highly saturated concentration of 2000 mg/dL. A stock of 2000 mg/dL was prepared for the spectrum measurements, and a stock of 400 mg/dL was considered for the optical rotation measurements. PA spectrum was acquired with 4 different concentrations i.e. 100 mg/dL, 200 mg/dL, 300 mg/dL, and 2000 mg/dL. The concentration of 400 mg/dL was diluted by adding equal parts of water dispensed such that all the concentrations within the physiological range were covered for optical rotation measurements. The solution volume used in the custom-designed holder was 55 mL. All the samples were prepared at least 2.5 hours before the experiments at room temperature. The BSA-based samples were filtered using a syringe filter with a pore size of 0.22 $\mu$m. The samples prepared were refrigerated in 4$^{\circ}$ C post experiments. Care was taken to bring up the temperature from 4$^{\circ}$ C to ambient temperature before the experiments were repeated. For the glucose validation measurements, a commercially available glucometer (Accu Chek Verio) was used. A volume of 20 $\mu$L of each sample was pipetted out and placed on a petri dish, which is similar to the volume of blood produced on a finger prick to inject onto the glucometer strip.\\

\subsection{\textit{Ex vivo} Sample preparation} 
Freshly procured chicken breast tissues were used for the \textit{ex-vivo} experiments. The tissue was sliced into very thin slices using a dissection blade. Slices of thickness 2 mm and 3.5 mm were dissected, as shown in Fig. 4(b) and were used for the experiments. The slices were placed at positions and were held by a fixture at the bottom of the holder. A solid partition was made out of glue. The fixtures were held in place by silica gel for proper bonding. The space between two slices was filled with glucose solution dissolved in BSA of 4 g/mL. Concentration measurements were carried out here by varying the solution concentration from 400 mg/dL to 90 mg/dL. The volume of sample used for these experiments was 15 mL. US gel was filled within the space between the chicken slice and the transducer after the alignment was ensured. The experiment using PAPEORS was repeated twice to evaluate the system and collect more data points (60 measurements for V, P, and R together).
\\

\subsection{Naproxen sample preparation} 
Pharmaceutical testing grade Naproxen from Merck (N8280) was used for the rotation experiments. The drug was dissolved in ethanol as the drug was not soluble in water. Therefore, a 70\% (V/V) ethanol was used as the solvent. A stock solution of 3.8 mg/mL was prepared, and lower concentrations were evaluated by diluting the stock with 70\%(V/V) ethanol. Consumption of Naproxen is prescribed in terms of dosages. The drug distribution in the body using standard dosages is about 0.1 mg/mL. The data acquired consists of 3 sets for each incidence. A total of 90 PA measurements were performed as part of the rotation experiments. PA spectrum were  acquired with Naproxen sample to identify the peak absorption wavelength.\\ 

\subsection{Statistical Analysis} 
Statistical studies were performed using the glucometer readings corresponding to the data from multiple sets of experiments. The estimated concentrations using PAPEORS and the glucometer readings were analysed using a Clarke's Error Grid Analysis (CEGA) approach for the glucose samples with multiple repetitions over large concentration ranges. The CEG was plotted by considering a total of 45 measurements repeated three times for the physiological range of 90 to 400 mg/dL and the reference readings were obtained from the glucometer (Accu check Verio) for the serum-based samples. The glucometer measurements for the aqueous samples were omitted as the readings displayed a substantial deviation (Ref. Fig. S6(c)) from the expected concentration levels, and the known prepared concentration was used as a reference. The mean and standard deviation were used while computing the optical rotation using PA measurements for both glucose and Naproxen samples.
\\

\backmatter

\bmhead{Supplementary information}

Supplementary information is available for this work.

\bmhead{Acknowledgments}
Jaya Prakash acknowledges the IISc Seed Grant. Swathi Padmanabhan acknowledges the Ministry of Education scholarship for graduate studies.


\section*{Declarations}
\begin{itemize}
\item Funding \\
No specific funding was received for this project.

\item Conflict of interest/Competing interests (check journal-specific guidelines for which heading to use) \\ 
The authors have submitted a complete patent application (App No: IN202341086473) based on this work, which is currently in review at the Indian Patent office. There is no other competing interest.
\item Ethics approval \\
No ethics approval was needed for this study since no animal or human study was performed
\item Availability of data and materials \\
All relevant data of this work will be made available upon request.

\item Code availability  \\
The code of this work will be made available upon request.

\item Authors' contributions \\
JP proposed the research direction, supervised the project, and participated in selecting the chiral materials and data analysis. SP was involved in designing and performing the experiments, and data analysis. SP and JP wrote the manuscript. All authors gave final consent to the manuscript.

\end{itemize}

\noindent


\bibliography{First_draft}


\begin{thebibliography}{52}
\ifx \bisbn   \undefined \def \bisbn  #1{ISBN #1}\fi
\ifx \binits  \undefined \def \binits#1{#1}\fi
\ifx \bauthor  \undefined \def \bauthor#1{#1}\fi
\ifx \batitle  \undefined \def \batitle#1{#1}\fi
\ifx \bjtitle  \undefined \def \bjtitle#1{#1}\fi
\ifx \bvolume  \undefined \def \bvolume#1{\textbf{#1}}\fi
\ifx \byear  \undefined \def \byear#1{#1}\fi
\ifx \bissue  \undefined \def \bissue#1{#1}\fi
\ifx \bfpage  \undefined \def \bfpage#1{#1}\fi
\ifx \blpage  \undefined \def \blpage #1{#1}\fi
\ifx \burl  \undefined \def \burl#1{\textsf{#1}}\fi
\ifx \doiurl  \undefined \def \doiurl#1{\url{https://doi.org/#1}}\fi
\ifx \betal  \undefined \def \betal{\textit{et al.}}\fi
\ifx \binstitute  \undefined \def \binstitute#1{#1}\fi
\ifx \binstitutionaled  \undefined \def \binstitutionaled#1{#1}\fi
\ifx \bctitle  \undefined \def \bctitle#1{#1}\fi
\ifx \beditor  \undefined \def \beditor#1{#1}\fi
\ifx \bpublisher  \undefined \def \bpublisher#1{#1}\fi
\ifx \bbtitle  \undefined \def \bbtitle#1{#1}\fi
\ifx \bedition  \undefined \def \bedition#1{#1}\fi
\ifx \bseriesno  \undefined \def \bseriesno#1{#1}\fi
\ifx \blocation  \undefined \def \blocation#1{#1}\fi
\ifx \bsertitle  \undefined \def \bsertitle#1{#1}\fi
\ifx \bsnm \undefined \def \bsnm#1{#1}\fi
\ifx \bsuffix \undefined \def \bsuffix#1{#1}\fi
\ifx \bparticle \undefined \def \bparticle#1{#1}\fi
\ifx \barticle \undefined \def \barticle#1{#1}\fi
\bibcommenthead
\ifx \bconfdate \undefined \def \bconfdate #1{#1}\fi
\ifx \botherref \undefined \def \botherref #1{#1}\fi
\ifx \url \undefined \def \url#1{\textsf{#1}}\fi
\ifx \bchapter \undefined \def \bchapter#1{#1}\fi
\ifx \bbook \undefined \def \bbook#1{#1}\fi
\ifx \bcomment \undefined \def \bcomment#1{#1}\fi
\ifx \oauthor \undefined \def \oauthor#1{#1}\fi
\ifx \citeauthoryear \undefined \def \citeauthoryear#1{#1}\fi
\ifx \endbibitem  \undefined \def \endbibitem {}\fi
\ifx \bconflocation  \undefined \def \bconflocation#1{#1}\fi
\ifx \arxivurl  \undefined \def \arxivurl#1{\textsf{#1}}\fi
\csname PreBibitemsHook\endcsname

\bibitem[\protect\citeauthoryear{Marth}{2008}]{marth2008unified}
\begin{barticle}
\bauthor{\bsnm{Marth}, \binits{J.D.}}:
\batitle{A unified vision of the building blocks of life}.
\bjtitle{Nature cell biology}
\bvolume{10}(\bissue{9}),
\bfpage{1015}--\blpage{1015}
(\byear{2008})
\end{barticle}
\endbibitem

\bibitem[\protect\citeauthoryear{Kelvin}{2021}]{kelvin2021molecular}
\begin{bbook}
\bauthor{\bsnm{Kelvin}, \binits{W.T.B.}}:
\bbtitle{The Molecular Tactics of a Crystal}.
\bpublisher{Good Press}, \blocation{???}
(\byear{2021}).
\burl{https://books.google.co.in/books?id=2vXXDwAAQBAJ}
\end{bbook}
\endbibitem

\bibitem[\protect\citeauthoryear{H~Brooks et~al.}{2011}]{h2011significance}
\begin{barticle}
\bauthor{\bsnm{H~Brooks}, \binits{W.}},
\bauthor{\bsnm{C~Guida}, \binits{W.}},
\bauthor{\bsnm{G~Daniel}, \binits{K.}}:
\batitle{The significance of chirality in drug design and development}.
\bjtitle{Current topics in medicinal chemistry}
\bvolume{11}(\bissue{7}),
\bfpage{760}--\blpage{770}
(\byear{2011})
\end{barticle}
\endbibitem

\bibitem[\protect\citeauthoryear{Kypr et~al.}{2009}]{kypr2009circular}
\begin{barticle}
\bauthor{\bsnm{Kypr}, \binits{J.}},
\bauthor{\bsnm{Kejnovsk{\'a}}, \binits{I.}},
\bauthor{\bsnm{Ren{\v{c}}iuk}, \binits{D.}},
\bauthor{\bsnm{Vorl{\'\i}{\v{c}}kov{\'a}}, \binits{M.}}:
\batitle{Circular dichroism and conformational polymorphism of dna}.
\bjtitle{Nucleic acids research}
\bvolume{37}(\bissue{6}),
\bfpage{1713}--\blpage{1725}
(\byear{2009})
\end{barticle}
\endbibitem

\bibitem[\protect\citeauthoryear{Morrow et~al.}{2017}]{morrow2017transmission}
\begin{barticle}
\bauthor{\bsnm{Morrow}, \binits{S.M.}},
\bauthor{\bsnm{Bissette}, \binits{A.J.}},
\bauthor{\bsnm{Fletcher}, \binits{S.P.}}:
\batitle{Transmission of chirality through space and across length scales}.
\bjtitle{Nature nanotechnology}
\bvolume{12}(\bissue{5}),
\bfpage{410}--\blpage{419}
(\byear{2017})
\end{barticle}
\endbibitem

\bibitem[\protect\citeauthoryear{B{\'e}gin et~al.}{2023}]{begin2023nonlinear}
\begin{barticle}
\bauthor{\bsnm{B{\'e}gin}, \binits{J.-L.}},
\bauthor{\bsnm{Jain}, \binits{A.}},
\bauthor{\bsnm{Parks}, \binits{A.}},
\bauthor{\bsnm{Hufnagel}, \binits{F.}},
\bauthor{\bsnm{Corkum}, \binits{P.}},
\bauthor{\bsnm{Karimi}, \binits{E.}},
\bauthor{\bsnm{Brabec}, \binits{T.}},
\bauthor{\bsnm{Bhardwaj}, \binits{R.}}:
\batitle{Nonlinear helical dichroism in chiral and achiral molecules}.
\bjtitle{Nature Photonics}
\bvolume{17}(\bissue{1}),
\bfpage{82}--\blpage{88}
(\byear{2023})
\end{barticle}
\endbibitem

\bibitem[\protect\citeauthoryear{Berova et~al.}{2007}]{berova2007application}
\begin{barticle}
\bauthor{\bsnm{Berova}, \binits{N.}},
\bauthor{\bsnm{Di~Bari}, \binits{L.}},
\bauthor{\bsnm{Pescitelli}, \binits{G.}}:
\batitle{Application of electronic circular dichroism in configurational and conformational analysis of organic compounds}.
\bjtitle{Chemical Society Reviews}
\bvolume{36}(\bissue{6}),
\bfpage{914}--\blpage{931}
(\byear{2007})
\end{barticle}
\endbibitem

\bibitem[\protect\citeauthoryear{Therapontos et~al.}{2009}]{therapontos2009thalidomide}
\begin{barticle}
\bauthor{\bsnm{Therapontos}, \binits{C.}},
\bauthor{\bsnm{Erskine}, \binits{L.}},
\bauthor{\bsnm{Gardner}, \binits{E.R.}},
\bauthor{\bsnm{Figg}, \binits{W.D.}},
\bauthor{\bsnm{Vargesson}, \binits{N.}}:
\batitle{Thalidomide induces limb defects by preventing angiogenic outgrowth during early limb formation}.
\bjtitle{Proceedings of the national academy of sciences}
\bvolume{106}(\bissue{21}),
\bfpage{8573}--\blpage{8578}
(\byear{2009})
\end{barticle}
\endbibitem

\bibitem[\protect\citeauthoryear{Burley and Lenz}{1962}]{BURLEY1962271}
\begin{barticle}
\bauthor{\bsnm{Burley}, \binits{D.M.}},
\bauthor{\bsnm{Lenz}, \binits{W.}}:
\batitle{Thalidomide and congenital abnormalities}.
\bjtitle{The Lancet}
\bvolume{279}(\bissue{7223}),
\bfpage{271}--\blpage{272}
(\byear{1962})
\doiurl{10.1016/S0140-6736(62)91217-5} .
\bcomment{Originally published as Volume 1, Issue 7223}
\end{barticle}
\endbibitem

\bibitem[\protect\citeauthoryear{Liu et~al.}{2023}]{liu2023detection}
\begin{barticle}
\bauthor{\bsnm{Liu}, \binits{Y.}},
\bauthor{\bsnm{Wu}, \binits{Z.}},
\bauthor{\bsnm{Armstrong}, \binits{D.W.}},
\bauthor{\bsnm{Wolosker}, \binits{H.}},
\bauthor{\bsnm{Zheng}, \binits{Y.}}:
\batitle{Detection and analysis of chiral molecules as disease biomarkers}.
\bjtitle{Nature Reviews Chemistry}
\bvolume{7}(\bissue{5}),
\bfpage{355}--\blpage{373}
(\byear{2023})
\end{barticle}
\endbibitem

\bibitem[\protect\citeauthoryear{Habartov{\'a} et~al.}{2018}]{habartova2018chiroptical}
\begin{barticle}
\bauthor{\bsnm{Habartov{\'a}}, \binits{L.}},
\bauthor{\bsnm{Bungani{\v{c}}}, \binits{B.}},
\bauthor{\bsnm{Tatarkovi{\v{c}}}, \binits{M.}},
\bauthor{\bsnm{Zavoral}, \binits{M.}},
\bauthor{\bsnm{Vondrou{\v{s}}ov{\'a}}, \binits{J.}},
\bauthor{\bsnm{Syslov{\'a}}, \binits{K.}},
\bauthor{\bsnm{Setni{\v{c}}ka}, \binits{V.}}:
\batitle{Chiroptical spectroscopy and metabolomics for blood-based sensing of pancreatic cancer}.
\bjtitle{Chirality}
\bvolume{30}(\bissue{5}),
\bfpage{581}--\blpage{591}
(\byear{2018})
\end{barticle}
\endbibitem

\bibitem[\protect\citeauthoryear{Wang et~al.}{2024}]{WANG2024115759}
\begin{barticle}
\bauthor{\bsnm{Wang}, \binits{X.}},
\bauthor{\bsnm{Chen}, \binits{J.}},
\bauthor{\bsnm{Xu}, \binits{H.}},
\bauthor{\bsnm{Fan}, \binits{Y.}},
\bauthor{\bsnm{Wang}, \binits{X.}},
\bauthor{\bsnm{Zhang}, \binits{M.}},
\bauthor{\bsnm{Liu}, \binits{Y.}},
\bauthor{\bsnm{Li}, \binits{B.}},
\bauthor{\bsnm{Liu}, \binits{J.}},
\bauthor{\bsnm{Zhou}, \binits{H.}}:
\batitle{Construction of an ultrasensitive dual-mode chiral molecules sensing platform based on molecularly imprinted polymer modified bipolar electrode}.
\bjtitle{Biosensors and Bioelectronics}
\bvolume{243},
\bfpage{115759}
(\byear{2024})
\doiurl{10.1016/j.bios.2023.115759}
\end{barticle}
\endbibitem

\bibitem[\protect\citeauthoryear{He et~al.}{2021}]{he2021polarisation}
\begin{barticle}
\bauthor{\bsnm{He}, \binits{C.}},
\bauthor{\bsnm{He}, \binits{H.}},
\bauthor{\bsnm{Chang}, \binits{J.}},
\bauthor{\bsnm{Chen}, \binits{B.}},
\bauthor{\bsnm{Ma}, \binits{H.}},
\bauthor{\bsnm{Booth}, \binits{M.J.}}:
\batitle{Polarisation optics for biomedical and clinical applications: a review}.
\bjtitle{Light: Science \& Applications}
\bvolume{10}(\bissue{1}),
\bfpage{194}
(\byear{2021})
\end{barticle}
\endbibitem

\bibitem[\protect\citeauthoryear{Ghosh and Vitkin}{2011}]{ghosh2011tissue}
\begin{barticle}
\bauthor{\bsnm{Ghosh}, \binits{N.}},
\bauthor{\bsnm{Vitkin}, \binits{I.A.}}:
\batitle{Tissue polarimetry: concepts, challenges, applications, and outlook}.
\bjtitle{Journal of biomedical optics}
\bvolume{16}(\bissue{11}),
\bfpage{110801}--\blpage{110801}
(\byear{2011})
\end{barticle}
\endbibitem

\bibitem[\protect\citeauthoryear{Lee et~al.}{2019}]{lee2019digital}
\begin{barticle}
\bauthor{\bsnm{Lee}, \binits{H.R.}},
\bauthor{\bsnm{Li}, \binits{P.}},
\bauthor{\bsnm{Yoo}, \binits{T.S.H.}},
\bauthor{\bsnm{Lotz}, \binits{C.}},
\bauthor{\bsnm{Groeber-Becker}, \binits{F.K.}},
\bauthor{\bsnm{Dembski}, \binits{S.}},
\bauthor{\bsnm{Garcia-Caurel}, \binits{E.}},
\bauthor{\bsnm{Ossikovski}, \binits{R.}},
\bauthor{\bsnm{Ma}, \binits{H.}},
\bauthor{\bsnm{Novikova}, \binits{T.}}:
\batitle{Digital histology with mueller microscopy: how to mitigate an impact of tissue cut thickness fluctuations}.
\bjtitle{Journal of Biomedical Optics}
\bvolume{24}(\bissue{7}),
\bfpage{076004}--\blpage{076004}
(\byear{2019})
\end{barticle}
\endbibitem

\bibitem[\protect\citeauthoryear{Wolfender et~al.}{2015}]{wolfender2015current}
\begin{barticle}
\bauthor{\bsnm{Wolfender}, \binits{J.-L.}},
\bauthor{\bsnm{Marti}, \binits{G.}},
\bauthor{\bsnm{Thomas}, \binits{A.}},
\bauthor{\bsnm{Bertrand}, \binits{S.}}:
\batitle{Current approaches and challenges for the metabolite profiling of complex natural extracts}.
\bjtitle{Journal of Chromatography A}
\bvolume{1382},
\bfpage{136}--\blpage{164}
(\byear{2015})
\end{barticle}
\endbibitem

\bibitem[\protect\citeauthoryear{Stark et~al.}{2019}]{stark2019broadband}
\begin{barticle}
\bauthor{\bsnm{Stark}, \binits{C.}},
\bauthor{\bsnm{Arrieta}, \binits{C.A.C.}},
\bauthor{\bsnm{Behroozian}, \binits{R.}},
\bauthor{\bsnm{Redmer}, \binits{B.}},
\bauthor{\bsnm{Fiedler}, \binits{F.}},
\bauthor{\bsnm{M{\"u}ller}, \binits{S.}}:
\batitle{Broadband polarimetric glucose determination in protein containing media using characteristic optical rotatory dispersion}.
\bjtitle{Biomedical Optics Express}
\bvolume{10}(\bissue{12}),
\bfpage{6340}--\blpage{6350}
(\byear{2019})
\end{barticle}
\endbibitem

\bibitem[\protect\citeauthoryear{Li et~al.}{2021}]{li2021measuring}
\begin{barticle}
\bauthor{\bsnm{Li}, \binits{D.}},
\bauthor{\bsnm{Xu}, \binits{C.}},
\bauthor{\bsnm{Zhang}, \binits{M.}},
\bauthor{\bsnm{Wang}, \binits{X.}},
\bauthor{\bsnm{Guo}, \binits{K.}},
\bauthor{\bsnm{Sun}, \binits{Y.}},
\bauthor{\bsnm{Gao}, \binits{J.}},
\bauthor{\bsnm{Guo}, \binits{Z.}}:
\batitle{Measuring glucose concentration in a solution based on the indices of polarimetric purity}.
\bjtitle{Biomedical Optics Express}
\bvolume{12}(\bissue{4}),
\bfpage{2447}--\blpage{2459}
(\byear{2021})
\end{barticle}
\endbibitem

\bibitem[\protect\citeauthoryear{Cameron et~al.}{1999}]{cameron1999use}
\begin{barticle}
\bauthor{\bsnm{Cameron}, \binits{B.D.}},
\bauthor{\bsnm{Gorde}, \binits{H.W.}},
\bauthor{\bsnm{Satheesan}, \binits{B.}},
\bauthor{\bsnm{Cote}, \binits{G.L.}}:
\batitle{The use of polarized laser light through the eye for noninvasive glucose monitoring}.
\bjtitle{Diabetes Technology \& Therapeutics}
\bvolume{1}(\bissue{2}),
\bfpage{135}--\blpage{143}
(\byear{1999})
\end{barticle}
\endbibitem

\bibitem[\protect\citeauthoryear{Jacques et~al.}{2002}]{jacques2002imaging}
\begin{barticle}
\bauthor{\bsnm{Jacques}, \binits{S.L.}},
\bauthor{\bsnm{Ramella-Roman}, \binits{J.C.}},
\bauthor{\bsnm{Lee}, \binits{K.}}:
\batitle{Imaging skin pathology with polarized light}.
\bjtitle{Journal of biomedical optics}
\bvolume{7}(\bissue{3}),
\bfpage{329}--\blpage{340}
(\byear{2002})
\end{barticle}
\endbibitem

\bibitem[\protect\citeauthoryear{Qi et~al.}{2023}]{qi2023surgical}
\begin{botherref}
\oauthor{\bsnm{Qi}, \binits{J.}},
\oauthor{\bsnm{Tatla}, \binits{T.}},
\oauthor{\bsnm{Nissanka-Jayasuriya}, \binits{E.}},
\oauthor{\bsnm{Yuan}, \binits{A.Y.}},
\oauthor{\bsnm{Stoyanov}, \binits{D.}},
\oauthor{\bsnm{Elson}, \binits{D.S.}}:
Surgical polarimetric endoscopy for the detection of laryngeal cancer.
Nature biomedical engineering,
1--15
(2023)
\end{botherref}
\endbibitem

\bibitem[\protect\citeauthoryear{Lippok et~al.}{2017}]{lippok2017depolarization}
\begin{barticle}
\bauthor{\bsnm{Lippok}, \binits{N.}},
\bauthor{\bsnm{Villiger}, \binits{M.}},
\bauthor{\bsnm{Albanese}, \binits{A.}},
\bauthor{\bsnm{Meijer}, \binits{E.F.}},
\bauthor{\bsnm{Chung}, \binits{K.}},
\bauthor{\bsnm{Padera}, \binits{T.P.}},
\bauthor{\bsnm{Bhatia}, \binits{S.N.}},
\bauthor{\bsnm{Bouma}, \binits{B.E.}}:
\batitle{Depolarization signatures map gold nanorods within biological tissue}.
\bjtitle{Nature photonics}
\bvolume{11}(\bissue{9}),
\bfpage{583}--\blpage{588}
(\byear{2017})
\end{barticle}
\endbibitem

\bibitem[\protect\citeauthoryear{Co\^{}~te{\'{}} and Vitkin}{2004}]{co2004balanced}
\begin{barticle}
\bauthor{\bsnm{Co\^{}~te{\'{}}}, \binits{D.}},
\bauthor{\bsnm{Vitkin}, \binits{I.A.}}:
\batitle{Balanced detection for low-noise precision polarimetric measurements of optically active, multiply scattering tissue phantoms}.
\bjtitle{Journal of biomedical optics}
\bvolume{9}(\bissue{1}),
\bfpage{213}--\blpage{220}
(\byear{2004})
\end{barticle}
\endbibitem

\bibitem[\protect\citeauthoryear{C{\^o}t{\'e} and Vitkin}{2005}]{cote2005robust}
\begin{barticle}
\bauthor{\bsnm{C{\^o}t{\'e}}, \binits{D.}},
\bauthor{\bsnm{Vitkin}, \binits{I.A.}}:
\batitle{Robust concentration determination of optically active molecules in turbid media with validated three-dimensional polarization sensitive monte carlo calculations}.
\bjtitle{Optics express}
\bvolume{13}(\bissue{1}),
\bfpage{148}--\blpage{163}
(\byear{2005})
\end{barticle}
\endbibitem

\bibitem[\protect\citeauthoryear{Greenfield}{2006}]{greenfield2006using}
\begin{barticle}
\bauthor{\bsnm{Greenfield}, \binits{N.J.}}:
\batitle{Using circular dichroism spectra to estimate protein secondary structure}.
\bjtitle{Nature protocols}
\bvolume{1}(\bissue{6}),
\bfpage{2876}--\blpage{2890}
(\byear{2006})
\end{barticle}
\endbibitem

\bibitem[\protect\citeauthoryear{Micsonai et~al.}{2015}]{micsonai2015accurate}
\begin{barticle}
\bauthor{\bsnm{Micsonai}, \binits{A.}},
\bauthor{\bsnm{Wien}, \binits{F.}},
\bauthor{\bsnm{Kernya}, \binits{L.}},
\bauthor{\bsnm{Lee}, \binits{Y.-H.}},
\bauthor{\bsnm{Goto}, \binits{Y.}},
\bauthor{\bsnm{R{\'e}fr{\'e}giers}, \binits{M.}},
\bauthor{\bsnm{Kardos}, \binits{J.}}:
\batitle{Accurate secondary structure prediction and fold recognition for circular dichroism spectroscopy}.
\bjtitle{Proceedings of the National Academy of Sciences}
\bvolume{112}(\bissue{24}),
\bfpage{3095}--\blpage{3103}
(\byear{2015})
\end{barticle}
\endbibitem

\bibitem[\protect\citeauthoryear{Kwon et~al.}{2023}]{kwon2023chiral}
\begin{botherref}
\oauthor{\bsnm{Kwon}, \binits{J.}},
\oauthor{\bsnm{Park}, \binits{K.H.}},
\oauthor{\bsnm{Choi}, \binits{W.J.}},
\oauthor{\bsnm{Kotov}, \binits{N.A.}},
\oauthor{\bsnm{Yeom}, \binits{J.}}:
Chiral spectroscopy of nanostructures.
Accounts of Chemical Research,
15229--15237
(2023)
\end{botherref}
\endbibitem

\bibitem[\protect\citeauthoryear{Arabi et~al.}{2022}]{arabi2022chiral}
\begin{barticle}
\bauthor{\bsnm{Arabi}, \binits{M.}},
\bauthor{\bsnm{Ostovan}, \binits{A.}},
\bauthor{\bsnm{Wang}, \binits{Y.}},
\bauthor{\bsnm{Mei}, \binits{R.}},
\bauthor{\bsnm{Fu}, \binits{L.}},
\bauthor{\bsnm{Li}, \binits{J.}},
\bauthor{\bsnm{Wang}, \binits{X.}},
\bauthor{\bsnm{Chen}, \binits{L.}}:
\batitle{Chiral molecular imprinting-based sers detection strategy for absolute enantiomeric discrimination}.
\bjtitle{Nature Communications}
\bvolume{13}(\bissue{1}),
\bfpage{5757}
(\byear{2022})
\end{barticle}
\endbibitem

\bibitem[\protect\citeauthoryear{Palomo et~al.}{2022}]{palomo2022simultaneous}
\begin{barticle}
\bauthor{\bsnm{Palomo}, \binits{L.}},
\bauthor{\bsnm{Favereau}, \binits{L.}},
\bauthor{\bsnm{Senthilkumar}, \binits{K.}},
\bauthor{\bsnm{St{\k{e}}pie{\'n}}, \binits{M.}},
\bauthor{\bsnm{Casado}, \binits{J.}},
\bauthor{\bsnm{Ram{\'\i}rez}, \binits{F.J.}}:
\batitle{Simultaneous detection of circularly polarized luminescence and raman optical activity in an organic molecular lemniscate}.
\bjtitle{Angewandte Chemie International Edition}
\bvolume{61}(\bissue{34}),
\bfpage{202206976}
(\byear{2022})
\end{barticle}
\endbibitem

\bibitem[\protect\citeauthoryear{Abdali and Blanch}{2008}]{abdali2008surface}
\begin{barticle}
\bauthor{\bsnm{Abdali}, \binits{S.}},
\bauthor{\bsnm{Blanch}, \binits{E.W.}}:
\batitle{Surface enhanced raman optical activity (seroa)}.
\bjtitle{Chemical Society Reviews}
\bvolume{37}(\bissue{5}),
\bfpage{980}--\blpage{992}
(\byear{2008})
\end{barticle}
\endbibitem

\bibitem[\protect\citeauthoryear{Haesler et~al.}{2007}]{haesler2007absolute}
\begin{barticle}
\bauthor{\bsnm{Haesler}, \binits{J.}},
\bauthor{\bsnm{Schindelholz}, \binits{I.}},
\bauthor{\bsnm{Riguet}, \binits{E.}},
\bauthor{\bsnm{Bochet}, \binits{C.G.}},
\bauthor{\bsnm{Hug}, \binits{W.}}:
\batitle{Absolute configuration of chirally deuterated neopentane}.
\bjtitle{Nature}
\bvolume{446}(\bissue{7135}),
\bfpage{526}--\blpage{529}
(\byear{2007})
\end{barticle}
\endbibitem

\bibitem[\protect\citeauthoryear{Parcha{\v{n}}sk{\`y} et~al.}{2014}]{parchavnsky2014inspecting}
\begin{barticle}
\bauthor{\bsnm{Parcha{\v{n}}sk{\`y}}, \binits{V.}},
\bauthor{\bsnm{Kapit{\'a}n}, \binits{J.}},
\bauthor{\bsnm{Bou{\v{r}}}, \binits{P.}}:
\batitle{Inspecting chiral molecules by raman optical activity spectroscopy}.
\bjtitle{RSC Advances}
\bvolume{4}(\bissue{100}),
\bfpage{57125}--\blpage{57136}
(\byear{2014})
\end{barticle}
\endbibitem

\bibitem[\protect\citeauthoryear{Taruttis and Ntziachristos}{2015}]{taruttis2015advances}
\begin{barticle}
\bauthor{\bsnm{Taruttis}, \binits{A.}},
\bauthor{\bsnm{Ntziachristos}, \binits{V.}}:
\batitle{Advances in real-time multispectral optoacoustic imaging and its applications}.
\bjtitle{Nature photonics}
\bvolume{9}(\bissue{4}),
\bfpage{219}--\blpage{227}
(\byear{2015})
\end{barticle}
\endbibitem

\bibitem[\protect\citeauthoryear{Tripathi et~al.}{2023}]{tripathi2023seed}
\begin{barticle}
\bauthor{\bsnm{Tripathi}, \binits{M.}},
\bauthor{\bsnm{Padmanabhan}, \binits{S.}},
\bauthor{\bsnm{Prakash}, \binits{J.}},
\bauthor{\bsnm{Raichur}, \binits{A.M.}}:
\batitle{Seed-mediated galvanic synthesis of cus--au nanohybrids for photo-theranostic applications}.
\bjtitle{ACS Applied Nano Materials}
\bvolume{6}(\bissue{16}),
\bfpage{14861}--\blpage{14875}
(\byear{2023})
\end{barticle}
\endbibitem

\bibitem[\protect\citeauthoryear{Liu et~al.}{2021}]{liu2021croconaine}
\begin{barticle}
\bauthor{\bsnm{Liu}, \binits{N.}},
\bauthor{\bsnm{Gujrati}, \binits{V.}},
\bauthor{\bsnm{Malekzadeh-Najafabadi}, \binits{J.}},
\bauthor{\bsnm{Werner}, \binits{J.P.F.}},
\bauthor{\bsnm{Klemm}, \binits{U.}},
\bauthor{\bsnm{Tang}, \binits{L.}},
\bauthor{\bsnm{Chen}, \binits{Z.}},
\bauthor{\bsnm{Prakash}, \binits{J.}},
\bauthor{\bsnm{Huang}, \binits{Y.}},
\bauthor{\bsnm{Stiel}, \binits{A.}}, \betal:
\batitle{Croconaine-based nanoparticles enable efficient optoacoustic imaging of murine brain tumors}.
\bjtitle{Photoacoustics}
\bvolume{22},
\bfpage{100263}
(\byear{2021})
\end{barticle}
\endbibitem

\bibitem[\protect\citeauthoryear{Hong et~al.}{2014}]{hong2014through}
\begin{barticle}
\bauthor{\bsnm{Hong}, \binits{G.}},
\bauthor{\bsnm{Diao}, \binits{S.}},
\bauthor{\bsnm{Chang}, \binits{J.}},
\bauthor{\bsnm{Antaris}, \binits{A.L.}},
\bauthor{\bsnm{Chen}, \binits{C.}},
\bauthor{\bsnm{Zhang}, \binits{B.}},
\bauthor{\bsnm{Zhao}, \binits{S.}},
\bauthor{\bsnm{Atochin}, \binits{D.N.}},
\bauthor{\bsnm{Huang}, \binits{P.L.}},
\bauthor{\bsnm{Andreasson}, \binits{K.I.}}, \betal:
\batitle{Through-skull fluorescence imaging of the brain in a new near-infrared window}.
\bjtitle{Nature photonics}
\bvolume{8}(\bissue{9}),
\bfpage{723}--\blpage{730}
(\byear{2014})
\end{barticle}
\endbibitem

\bibitem[\protect\citeauthoryear{Miao and Pu}{2018}]{miao2018organic}
\begin{barticle}
\bauthor{\bsnm{Miao}, \binits{Q.}},
\bauthor{\bsnm{Pu}, \binits{K.}}:
\batitle{Organic semiconducting agents for deep-tissue molecular imaging: second near-infrared fluorescence, self-luminescence, and photoacoustics}.
\bjtitle{Advanced materials}
\bvolume{30}(\bissue{49}),
\bfpage{1801778}
(\byear{2018})
\end{barticle}
\endbibitem

\bibitem[\protect\citeauthoryear{Hong et~al.}{2017}]{hong2017near}
\begin{barticle}
\bauthor{\bsnm{Hong}, \binits{G.}},
\bauthor{\bsnm{Antaris}, \binits{A.L.}},
\bauthor{\bsnm{Dai}, \binits{H.}}:
\batitle{Near-infrared fluorophores for biomedical imaging}.
\bjtitle{Nature biomedical engineering}
\bvolume{1}(\bissue{1}),
\bfpage{0010}
(\byear{2017})
\end{barticle}
\endbibitem

\bibitem[\protect\citeauthoryear{Shi et~al.}{2019}]{shi2019high}
\begin{barticle}
\bauthor{\bsnm{Shi}, \binits{J.}},
\bauthor{\bsnm{Wong}, \binits{T.T.}},
\bauthor{\bsnm{He}, \binits{Y.}},
\bauthor{\bsnm{Li}, \binits{L.}},
\bauthor{\bsnm{Zhang}, \binits{R.}},
\bauthor{\bsnm{Yung}, \binits{C.S.}},
\bauthor{\bsnm{Hwang}, \binits{J.}},
\bauthor{\bsnm{Maslov}, \binits{K.}},
\bauthor{\bsnm{Wang}, \binits{L.V.}}:
\batitle{High-resolution, high-contrast mid-infrared imaging of fresh biological samples with ultraviolet-localized photoacoustic microscopy}.
\bjtitle{Nature photonics}
\bvolume{13}(\bissue{9}),
\bfpage{609}--\blpage{615}
(\byear{2019})
\end{barticle}
\endbibitem

\bibitem[\protect\citeauthoryear{Prakash et~al.}{2020}]{prakash2020short}
\begin{barticle}
\bauthor{\bsnm{Prakash}, \binits{J.}},
\bauthor{\bsnm{Seyedebrahimi}, \binits{M.M.}},
\bauthor{\bsnm{Ghazaryan}, \binits{A.}},
\bauthor{\bsnm{Malekzadeh-Najafabadi}, \binits{J.}},
\bauthor{\bsnm{Gujrati}, \binits{V.}},
\bauthor{\bsnm{Ntziachristos}, \binits{V.}}:
\batitle{Short-wavelength optoacoustic spectroscopy based on water muting}.
\bjtitle{Proceedings of the National Academy of Sciences}
\bvolume{117}(\bissue{8}),
\bfpage{4007}--\blpage{4014}
(\byear{2020})
\end{barticle}
\endbibitem

\bibitem[\protect\citeauthoryear{Ghazaryan et~al.}{2018}]{ghazaryan2018extended}
\begin{barticle}
\bauthor{\bsnm{Ghazaryan}, \binits{A.}},
\bauthor{\bsnm{Ovsepian}, \binits{S.V.}},
\bauthor{\bsnm{Ntziachristos}, \binits{V.}}:
\batitle{Extended near-infrared optoacoustic spectrometry for sensing physiological concentrations of glucose}.
\bjtitle{Frontiers in endocrinology}
\bvolume{9},
\bfpage{112}
(\byear{2018})
\end{barticle}
\endbibitem

\bibitem[\protect\citeauthoryear{Qu et~al.}{2018}]{qu2018dichroism}
\begin{barticle}
\bauthor{\bsnm{Qu}, \binits{Y.}},
\bauthor{\bsnm{Li}, \binits{L.}},
\bauthor{\bsnm{Shen}, \binits{Y.}},
\bauthor{\bsnm{Wei}, \binits{X.}},
\bauthor{\bsnm{Wong}, \binits{T.T.}},
\bauthor{\bsnm{Hu}, \binits{P.}},
\bauthor{\bsnm{Yao}, \binits{J.}},
\bauthor{\bsnm{Maslov}, \binits{K.}},
\bauthor{\bsnm{Wang}, \binits{L.V.}}:
\batitle{Dichroism-sensitive photoacoustic computed tomography}.
\bjtitle{Optica}
\bvolume{5}(\bissue{4}),
\bfpage{495}--\blpage{501}
(\byear{2018})
\end{barticle}
\endbibitem

\bibitem[\protect\citeauthoryear{Zhou et~al.}{2019}]{zhou2019single}
\begin{barticle}
\bauthor{\bsnm{Zhou}, \binits{Y.}},
\bauthor{\bsnm{Chen}, \binits{J.}},
\bauthor{\bsnm{Liu}, \binits{C.}},
\bauthor{\bsnm{Liu}, \binits{C.}},
\bauthor{\bsnm{Lai}, \binits{P.}},
\bauthor{\bsnm{Wang}, \binits{L.}}:
\batitle{Single-shot linear dichroism optical-resolution photoacoustic microscopy}.
\bjtitle{Photoacoustics}
\bvolume{16},
\bfpage{100148}
(\byear{2019})
\end{barticle}
\endbibitem

\bibitem[\protect\citeauthoryear{Zhang et~al.}{2023}]{zhang2023collagen}
\begin{barticle}
\bauthor{\bsnm{Zhang}, \binits{Z.}},
\bauthor{\bsnm{Chen}, \binits{W.}},
\bauthor{\bsnm{Cui}, \binits{D.}},
\bauthor{\bsnm{Mi}, \binits{J.}},
\bauthor{\bsnm{Mu}, \binits{G.}},
\bauthor{\bsnm{Nie}, \binits{L.}},
\bauthor{\bsnm{Yang}, \binits{S.}},
\bauthor{\bsnm{Shi}, \binits{Y.}}:
\batitle{Collagen fiber anisotropy characterization by polarized photoacoustic imaging for just-in-time quantitative evaluation of burn severity}.
\bjtitle{Photonics Research}
\bvolume{11}(\bissue{5}),
\bfpage{817}--\blpage{828}
(\byear{2023})
\end{barticle}
\endbibitem

\bibitem[\protect\citeauthoryear{Lide}{2004}]{lide2004crc}
\begin{bbook}
\bauthor{\bsnm{Lide}, \binits{D.R.}}:
\bbtitle{CRC Handbook of Chemistry and Physics}
vol. \bseriesno{85}.
\bpublisher{CRC press}, \blocation{???}
(\byear{2004})
\end{bbook}
\endbibitem

\bibitem[\protect\citeauthoryear{Barron}{2009}]{barron2009molecular}
\begin{bbook}
\bauthor{\bsnm{Barron}, \binits{L.D.}}:
\bbtitle{Molecular Light Scattering and Optical Activity}.
\bpublisher{Cambridge University Press}, \blocation{???}
(\byear{2009})
\end{bbook}
\endbibitem

\bibitem[\protect\citeauthoryear{Ramella-Roman et~al.}{2005}]{ramella2005three}
\begin{barticle}
\bauthor{\bsnm{Ramella-Roman}, \binits{J.C.}},
\bauthor{\bsnm{Prahl}, \binits{S.A.}},
\bauthor{\bsnm{Jacques}, \binits{S.L.}}:
\batitle{Three monte carlo programs of polarized light transport into scattering media: part i}.
\bjtitle{Optics Express}
\bvolume{13}(\bissue{12}),
\bfpage{4420}--\blpage{4438}
(\byear{2005})
\end{barticle}
\endbibitem

\bibitem[\protect\citeauthoryear{O'Neil}{2013}]{o2013merck}
\begin{bbook}
\bauthor{\bsnm{O'Neil}, \binits{M.J.}}:
\bbtitle{The Merck Index: an Encyclopedia of Chemicals, Drugs, and Biologicals}.
\bpublisher{RSC Publishing}, \blocation{???}
(\byear{2013})
\end{bbook}
\endbibitem

\bibitem[\protect\citeauthoryear{Li and Chen}{2023}]{li2023review}
\begin{botherref}
\oauthor{\bsnm{Li}, \binits{Y.}},
\oauthor{\bsnm{Chen}, \binits{Y.}}:
Review of noninvasive continuous glucose monitoring in diabetics.
ACS sensors
(2023)
\end{botherref}
\endbibitem

\bibitem[\protect\citeauthoryear{Karlas et~al.}{2023}]{karlas2023dermal}
\begin{botherref}
\oauthor{\bsnm{Karlas}, \binits{A.}},
\oauthor{\bsnm{Katsouli}, \binits{N.}},
\oauthor{\bsnm{Fasoula}, \binits{N.-A.}},
\oauthor{\bsnm{Bariotakis}, \binits{M.}},
\oauthor{\bsnm{Chlis}, \binits{N.-K.}},
\oauthor{\bsnm{Omar}, \binits{M.}},
\oauthor{\bsnm{He}, \binits{H.}},
\oauthor{\bsnm{Iakovakis}, \binits{D.}},
\oauthor{\bsnm{Sch{\"a}ffer}, \binits{C.}},
\oauthor{\bsnm{Kallmayer}, \binits{M.}}, et al.:
Dermal features derived from optoacoustic tomograms via machine learning correlate microangiopathy phenotypes with diabetes stage.
Nature Biomedical Engineering,
1--16
(2023)
\end{botherref}
\endbibitem

\bibitem[\protect\citeauthoryear{Westerveld et~al.}{2021}]{westerveld2021sensitive}
\begin{barticle}
\bauthor{\bsnm{Westerveld}, \binits{W.J.}},
\bauthor{\bsnm{Mahmud-Ul-Hasan}, \binits{M.}},
\bauthor{\bsnm{Shnaiderman}, \binits{R.}},
\bauthor{\bsnm{Ntziachristos}, \binits{V.}},
\bauthor{\bsnm{Rottenberg}, \binits{X.}},
\bauthor{\bsnm{Severi}, \binits{S.}},
\bauthor{\bsnm{Rochus}, \binits{V.}}:
\batitle{Sensitive, small, broadband and scalable optomechanical ultrasound sensor in silicon photonics}.
\bjtitle{Nature Photonics}
\bvolume{15}(\bissue{5}),
\bfpage{341}--\blpage{345}
(\byear{2021})
\end{barticle}
\endbibitem

\bibitem[\protect\citeauthoryear{Liang et~al.}{2022}]{liang2022optical}
\begin{barticle}
\bauthor{\bsnm{Liang}, \binits{Y.}},
\bauthor{\bsnm{Fu}, \binits{W.}},
\bauthor{\bsnm{Li}, \binits{Q.}},
\bauthor{\bsnm{Chen}, \binits{X.}},
\bauthor{\bsnm{Sun}, \binits{H.}},
\bauthor{\bsnm{Wang}, \binits{L.}},
\bauthor{\bsnm{Jin}, \binits{L.}},
\bauthor{\bsnm{Huang}, \binits{W.}},
\bauthor{\bsnm{Guan}, \binits{B.-O.}}:
\batitle{Optical-resolution functional gastrointestinal photoacoustic endoscopy based on optical heterodyne detection of ultrasound}.
\bjtitle{Nature Communications}
\bvolume{13}(\bissue{1}),
\bfpage{7604}
(\byear{2022})
\end{barticle}
\endbibitem

\end{thebibliography}

\end{document}


\renewcommand\thefigure{S\arabic{figure}}
\renewcommand\thetable{S\arabic{table}}

\title[Article Title]{Deep Tissue Sensing of Chiral Molecules using Polarization Enhanced Photoacoustics \\

Supplementary Information}


\author[1]{\fnm{Swathi} \sur{Padmanabhan}}\email{swathip@iisc.ac.in}

\author*[1]{\fnm{Jaya} \sur{Prakash}}\email{jayap@iisc.ac.in}

\affil*[1]{\orgdiv{Instrumentation and Applied Physics}, \orgname{Indian Institute of Science}, \orgaddress{\street{C V Raman Road}, \city{Bangalore}, \postcode{560012}, \state{Karnataka}, \country{India}}}



\keywords{Optical rotation, Photoacoustic sensing, Polarization, Chirality}



\maketitle
\tableofcontents

\newpage
\renewcommand{\thesection}{\large \Roman{section}} 

\section{\large Conventional and Chiroptical Methods: \textit{In-vivo} Feasibility Summary}
A summary of the conventional and chiroptical methods highlighting various parameters like penetration depth, sample type, sample quantity and \textit{in-vivo} feasibility in biosensing is shown in Table-S1. The various chiroptical properties that are summarised are - \textbf{CD}: Circular Dichroism, \textbf{ROA}: Raman Optical Activity, \textbf{SERS}: Surface Enhanced Raman Spectroscopy, \textbf{SEROA}: Surface Enhanced Raman Optical Activity, \textbf{SECD}: Surface Enhanced Circular Dichroism, \textbf{VCD}: Vibrational Circular Dichroism. 
\begin{table}
\centering
\begin{adjustbox}{angle=90}
\begin{tabular}{|c|c|c|c|c|c|}
\hline \textbf{Method}  & \textbf{Sample type} & \textbf{Sample Quantity} & \textbf{Penetration depth} & \begin{tabular}{l} 
\textbf{\textit{In- vivo}} \\
\textbf{feasibility}
\end{tabular} & \textbf{Cost} \\
\hline \textbf{Chromatography}\cite{liu2023detection} & \begin{tabular}{l} 
Tissue, Serum, \\
plasma, \\
urine, CSF, saliva,
\end{tabular} & $10 \mu \mathrm{l}-100 \mathrm{ml}$ & NA & $X$ & Medium \\
\hline \textbf{Electrophoresis}\cite{liu2023detection} &  \begin{tabular}{l} 
Serum, plasma, \\
urine, tissue
\end{tabular} & $100 \mathrm{nl}-1 \mathrm{ml}$ & NA & $X$ & Low \\
\hline \textbf{Enzyme assays}\cite{liu2023detection}  & \begin{tabular}{l} 
Serum, CSF, \\
plasma, urine, \\
saliva, tissue
\end{tabular} & $10 \mu \mathrm{l}-10 \mathrm{ml}$ & NA & $X$ & Low \\
\hline \textbf{NMR Spectroscopy}\cite{liu2023detection}  & Urine/Fluid & $10 \mu \mathrm{l}-100 \mathrm{ml}$ & NA & $X$ & High \\
\hline \textbf{Ultrasound }\cite{zhang2022specific,xiong2020ultrasound} &\begin{tabular}{l} 
Tissue*/ \\
Blood/Fluid
\end{tabular} & $2 \mathrm{~mL}$ & $0.1 \mathrm{~cm}$ & $X^{\textcolor{red}{*}}$ & Low\\
\hline \textbf{Polarimetry}\cite{liu2023detection} & \begin{tabular}{l} 
Tissue/ \\
Blood/Fluid
\end{tabular} & $10 \mathrm{ml}$ & \begin{tabular}{l} 
Limited penetration due \\
to multiple scattering \\
- Eye [Retina]\cite{ozdek2002assessment}: Upto 60 to \\
100um \\
- Skin (VIS)\cite{jacques2002imaging} $\sim$ up to $0.4 \mathrm{~mm}$ \\
- Skin (NIR)\cite{jacques1996polarized} $>$ 300um to \\
$1.5 \mathrm{~mm}$
\end{tabular} & $\sqrt{ }$ & Low \\
\hline \textbf{CD Spectroscopy} & CSF, tissue/Fluids & $100 \mu \mathrm{l}-100 \mathrm{ml}$ & NA & $X$ & Low \\
\hline \textbf{ROA Spectroscopy} & Fluid & $100 \mu \mathrm{l}-100 \mathrm{ml}$ & NA & $X$ & Medium \\
\hline \textbf{VCD Spectroscopy} & Fluid & $100 \mu \mathrm{l}-100 \mathrm{ml}$ & NA & $X$ & Medium \\
\hline \textbf{SERS/SEROA Spectroscopy} & Fluid & $10 \mu \mathrm{l}-1 \mathrm{ml}$ & $N A^{\textcolor{red}{*}}$ & $X^{\textcolor{red}{*}}$ & Low \\
\hline \textbf{SECD Spectroscopy}  & \begin{tabular}{l} 
Tissue, \\
Urine/Fluids
\end{tabular} & $10 \mu \mathrm{l}-100 \mu \mathrm{l}$ & $\mathrm{NA}^{\textcolor{red}{*}}$ & $X^{\textcolor{red}{*}}$ & Low \\
\hline \textbf{PAPEORS} & \begin{tabular}{l} 
Tissue/blood/Fluid
\end{tabular} & $15 \mathrm{~mL}^{\textcolor{red}{**}}$ & $ 3.5 \mathrm{~mm}$ & $\sqrt{ }$ & Scalable \\
\hline
\end{tabular}

\end{adjustbox}
\caption{\textbf{Summary of Chiral Sensing Techniques}: The table summarises the various parameters considered for chiral sensing like penetration depth and \textit{in-vivo} feasibility. Remarks:  X\textcolor{red}{$^*$} denotes that the system is under development and the method has potential for in-vivo experiments. $^{\textcolor{red}{**}}$ for PAPEORS denotes the current parameters, which can be improved by miniaturizing the system. The chiroptical properties that are summarized are - \textbf{CD}: Circular Dichroism, \textbf{ROA}: Raman Optical Activity, \textbf{SERS}: Surface Enhanced Raman Spectroscopy, \textbf{SEROA}: Surface Enhanced Raman Optical Activity, \textbf{SECD}: Surface Enhanced Circular Dichroism, \textbf{VCD}: Vibrational Circular Dichroism.  }
    \label{tab:my_label}
\end{table}

\section{\large PAPEORS Principle}
The recorded photoacoustic signal [$y(t)$] was observed to be a convolved output of the transducer impulse response [$h(t)$] and the original PA signal [$p(t)$],
\begin{equation}
         y(t,\phi, \mu_a) = p(t,\phi, \mu_a)*h(t)
\end{equation}
Deconvolution of the recorded signal with the transducer impulse response was performed to determine the $P$ and $P_0$ amplitudes accurately. The obtained signals were deconvolved in the frequency domain to recover the photoacoustic signal as a function of depth. The Fourier transform of the Eq. 1 results in,
 \begin{equation}
         Y(\omega) = P(\omega)H(\omega)
\end{equation}
The deconvolved PA signal in the frequency domain can be written as,
\begin{equation}
         P(\omega) = \dfrac{Y(\omega)}{H(\omega)}
\end{equation}
The time-series photoacoustic data is then correlated to the depth for further extrapolation of the PA amplitudes at various depths.
\begin{equation}
         p(t,\phi, \mu_a) = p(d,\phi, \mu_a)
     \end{equation}
where $d$ is the depth or path length. 
The detailed steps involved in extracting the points after deconvolution from $ p(d,\phi, \mu_a)$ are illustrated in Fig. S1.
\begin{figure}
\centering
\includegraphics[width=1\linewidth]{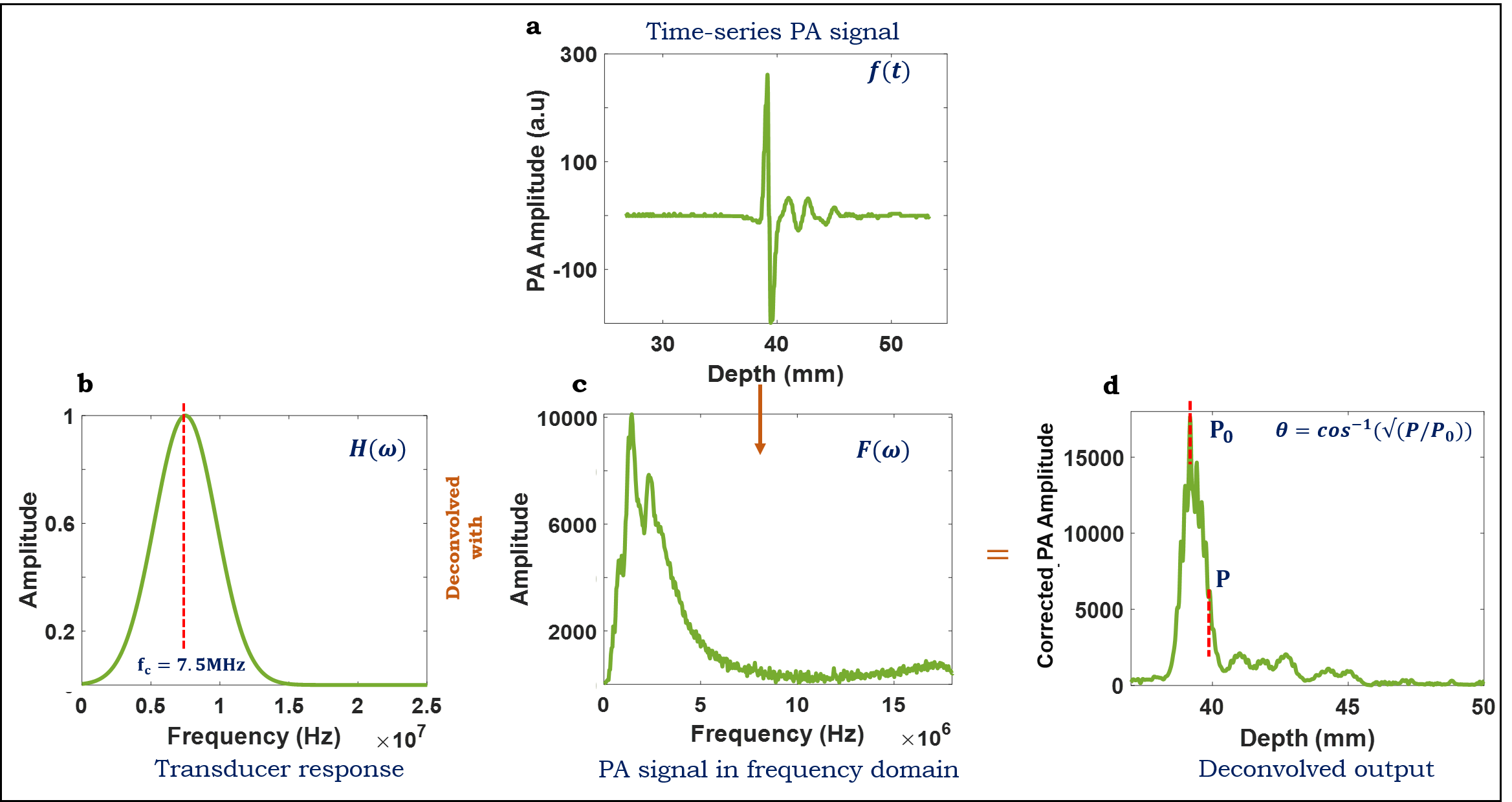}

\caption{\textbf{Deconvolution of the time-series photoacoustic signal:} \textbf{(a)} shows the raw PA time-series signal, $f(t)$ from the experiment, which is a convolved output of the transducer response and the original PA signal. \textbf{(b)} is the transducer response ($H(\omega)$) of the transducer used for the acquisition with the center frequency $f_c = 7.5MHz$. The response is reproduced from the datasheet provided by Olympus. \textbf{(c)} shows the frequency domain representation of the signal acquired, $F(\omega)$. \textbf{(d)} shows the final deconvolved output after inverse Fourier Transform, which is the original PA signal that is later used for computing rotation and further calculations.     }
\label{<figure-label>}
\end{figure}

\begin{figure}
    \centering
    \includegraphics[width=1\linewidth]{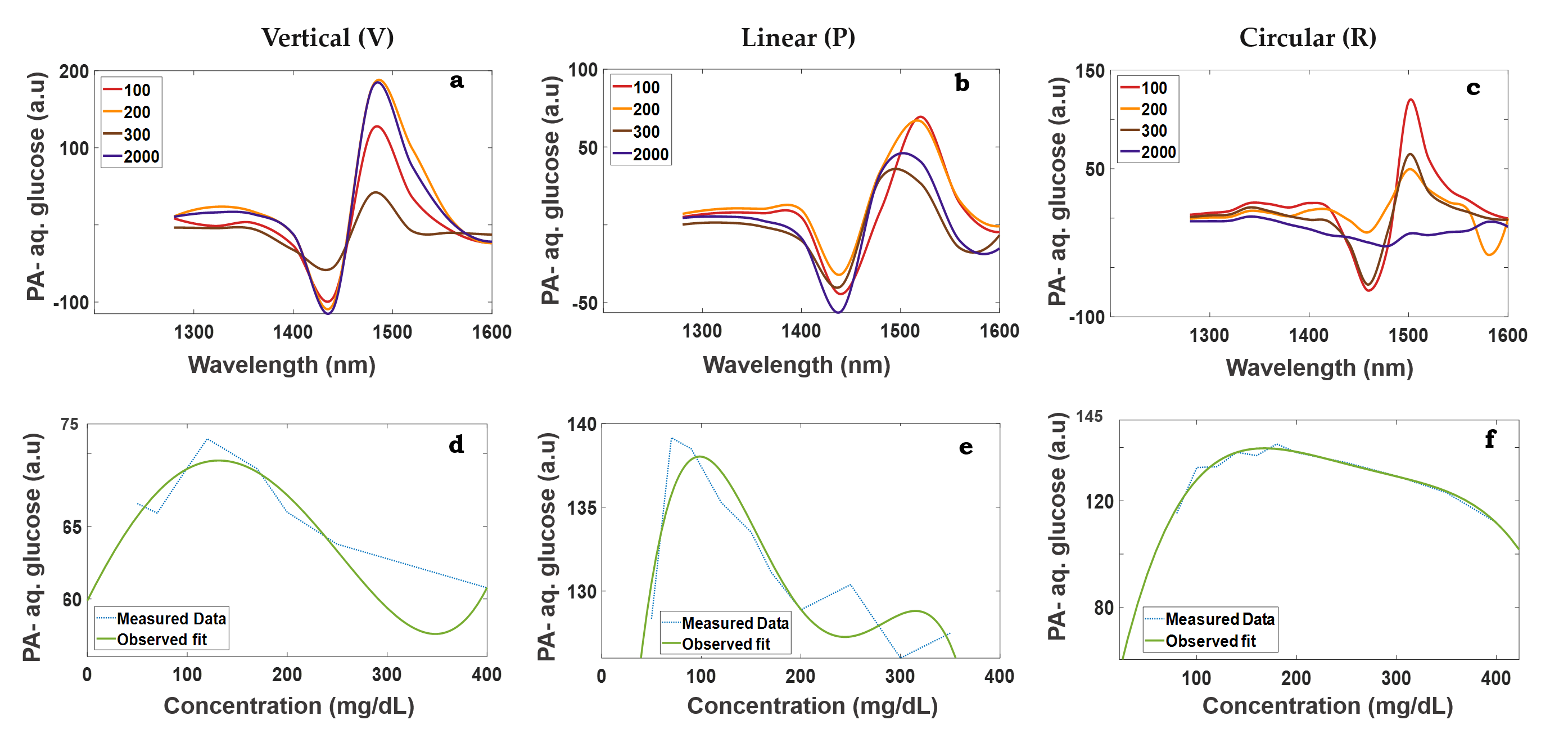}
    \caption{\textbf{Photoacoustic spectrum and non-linear effects with aqueous glucose samples} PA Spectra for \textbf{(a)} Vertical (V) incidence, \textbf{(b)} Linear (P) incidence, and \textbf{(c),}  for the Circular (R) incidence.  \textbf{(d), (e) } and \textbf{(f)} show the non-linear variation of the PA amplitude as a function of concentration at a depth of 1.7 mm for V, P, and R incidences, respectively.} 
    \label{fig:enter-label}
\end{figure}

\begin{figure}
\centering
\includegraphics[width=1\linewidth]{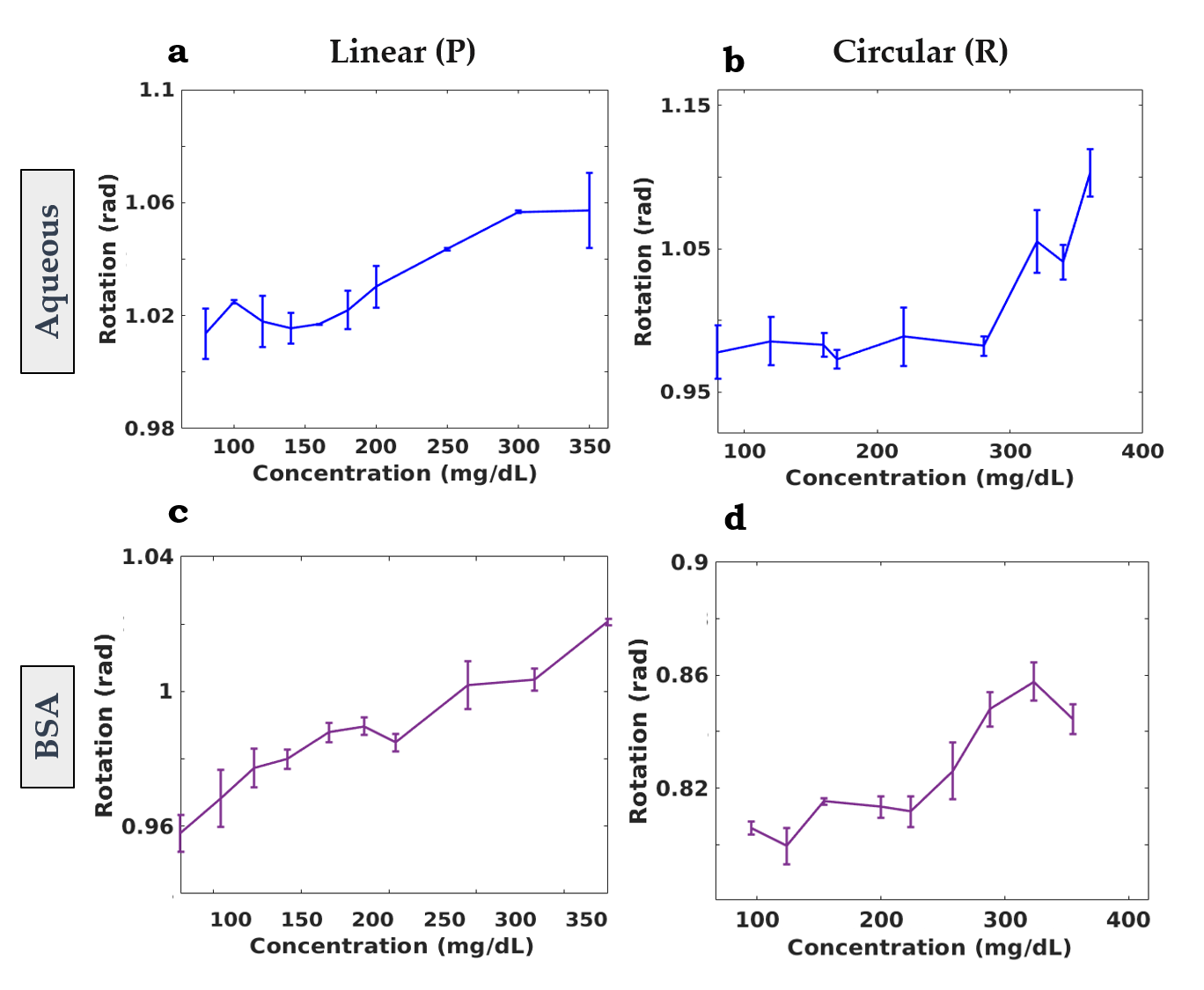}
\rule{0pt}{6ex}
\caption{\textbf{Rotation from PA experiments with glucose samples:} \textbf{Top row}- Aqueous glucose samples and \textbf{Bottom row} BSA-based glucose samples. \textbf{(a)-(b)} represent the variation in rotation as a function of concentration at 1.7 mm for P and R polarized incidence, respectively. \textbf{(c)-(d)} represent the rotation as a function of concentration for BSA glucose samples at the same depth. The analysis showed that the variance in BSA glucose samples is lower than in the aqueous samples. The trends observed in these plots formed the basis for building the concentration estimation model. The concentrations evaluated were in the physiological range of blood glucose levels: 50-400mg/dL and BSA concentration 4g/mL of serum albumin range.}
\label{<figure-label>}
\end{figure}

\renewcommand{\thesection}{\large \Roman{section}} 

\section{\large  Validation with Monte Carlo Simulations.}
Polarised Monte Carlo algorithm\cite{ramella2005three} was used to simulate the fluence profile across the thickness of the sample considered. The near-infrared wavelengths were used for the experiment. The optical properties were similar to the experimental conditions for aqueous glucose samples were simulated to calculate the optical rotation. The absorption coefficient corresponding to different concentrations was calculated based on the molar absorptivity\cite{amerov2004molar} at 1560 nm, was then used in the simulations. Fig. S4 shows the validation with the rotation from the experiments. 
\begin{figure}
    \centering
    \includegraphics[width=1\linewidth]{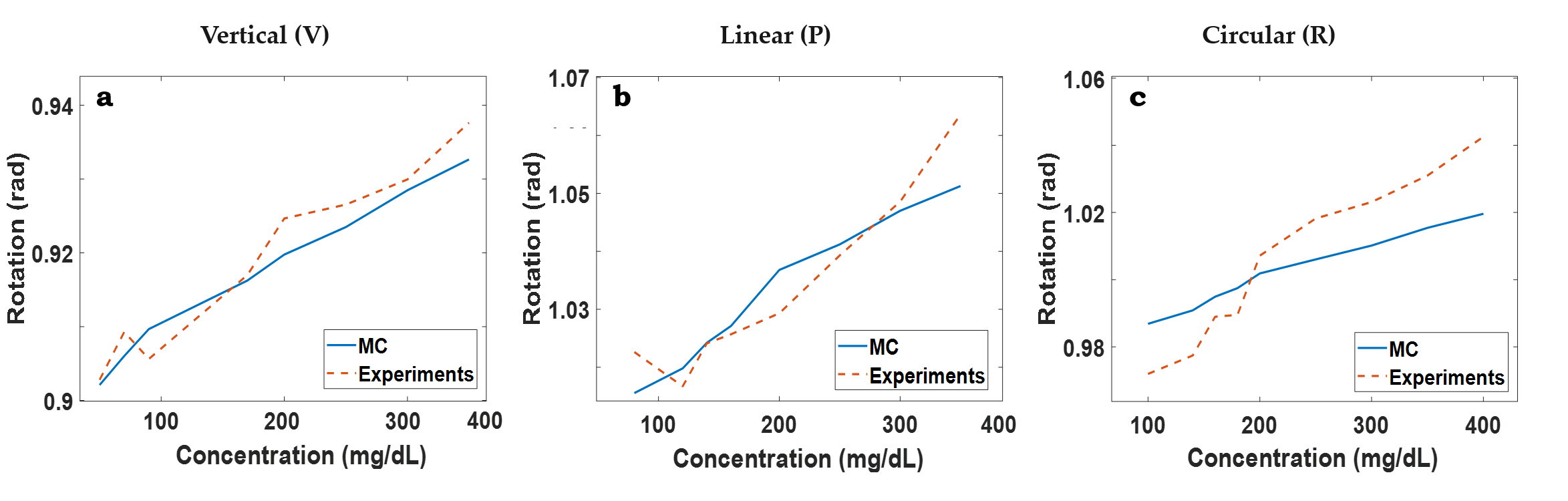}
    \caption{\textbf{Validation with Monte Carlo Simulations:} The rotation extrapolated from the fluence profiles utilising the Polarized Monte Carlo, considering a depth of 1.7mm. The absorption coefficient corresponding to different concentrations was calculated based on the molar absorptivity at 1560nm for the simulations. The scattering coefficient was considered to be low($\mu_s = 2 cm^{-1}$) considering the wavelength of incidence. The simulations were correlated with the optical rotation obtained from the fluence profiles using PA measurements. The figure shows the validation for V, P, and R incidences, respectively.}
    \label{fig:enter-label}
\end{figure}

\begin{figure}
    \centering
    \includegraphics[width=1\linewidth]{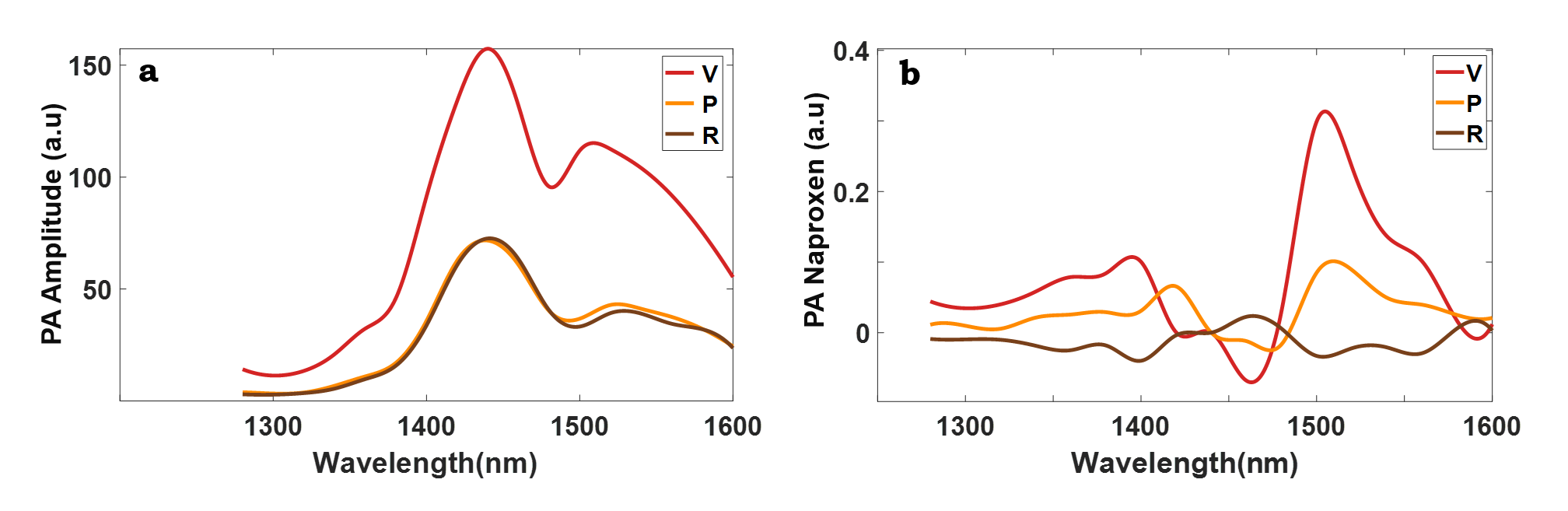}
    \caption{\textbf{Photoacoustic Spectra for Naproxen: (a)} shows the PA spectra for the Naproxen in 70\% ethanol for the V, P and R incidences. The dominant absorption of water and ethanol can be observed in \textbf{(a)}. The spectra after correcting for the ethanol absorption can be seen in \textbf{(b)}. The unique absorption of Naproxen was identified to be close to 1500 nm and was further used for the rotation experiments. Changes was observed in the Naproxen spectra with mild wavelength shift in the peak absorption. V and P incidence shows a similar trend, while R incident configuration does not show a clear peak in comparison to V and P. } 
    \label{fig:enter-label}
\end{figure}

\begin{figure}
    \centering
    \includegraphics[width=1\linewidth]{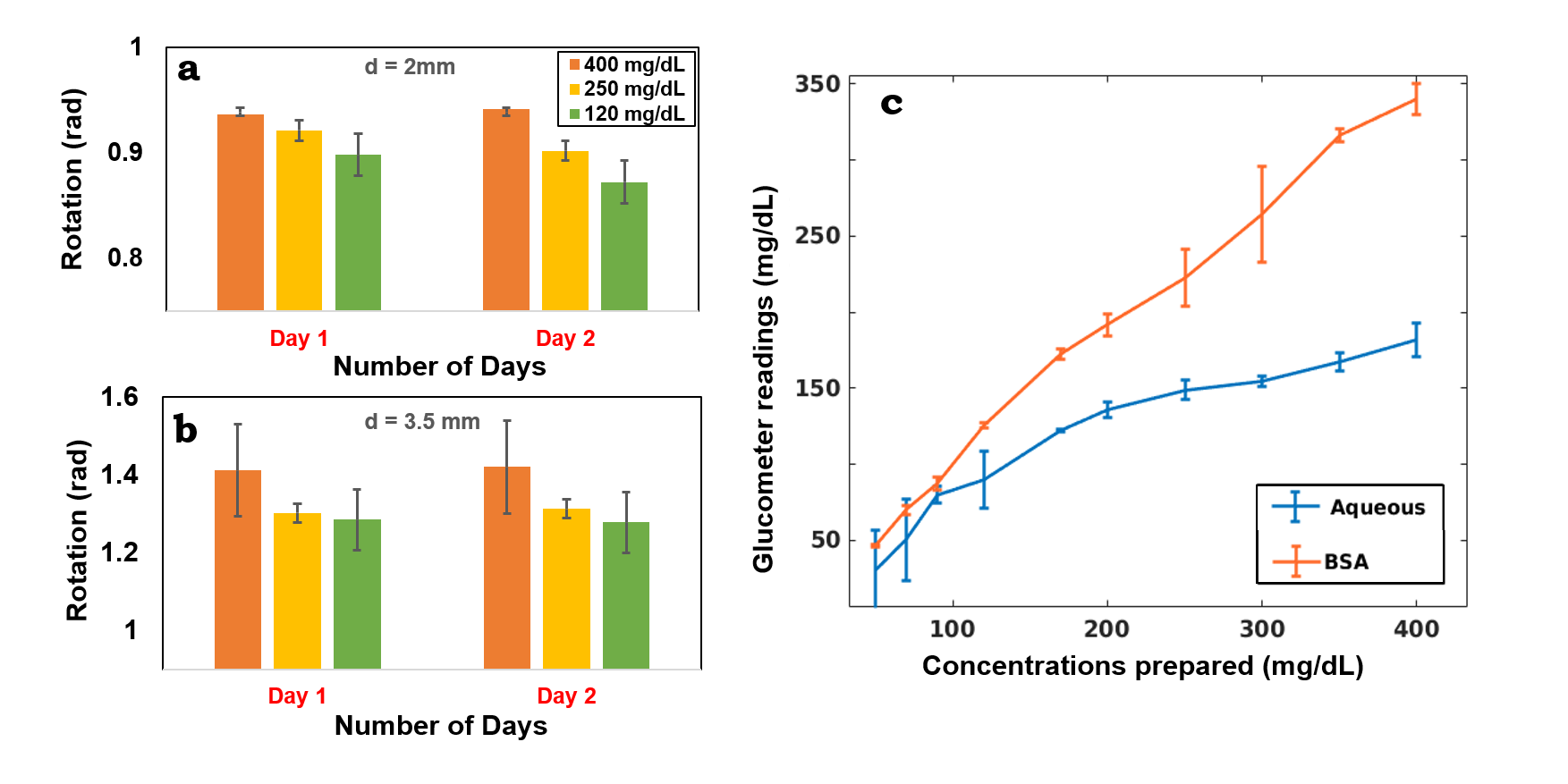}
    \caption{\textbf{Repeatability and Glucometer Validation} \textbf{(a)} and \textbf{(b)} shows the repeatability of extrapolating the magnitude of rotation from the experiments from two different days for 400 mg/dL, 250 mg/dL and 120 mg/dL for V incidence for glucose samples. \textbf{(a)} shows the data for a depth of 2mm and \textbf{(b)} for 3.5mm. The standard deviation is shown for Day 1 and Day 2 as inset. \textbf{(c)} The concentrations prepared are contrasted with the glucometer readings for the aqueous and BSA glucose samples. The deviation for the aqueous glucose samples is found to be quite high compared to the reading obtained from the BSA glucose samples. The possible reason for that could be that the albumin present in the sample makes the sample more similar to the blood plasma.  }
    \label{fig:enter-label}
\end{figure}

\begin{figure}
    \centering
    \includegraphics[width=1\linewidth]{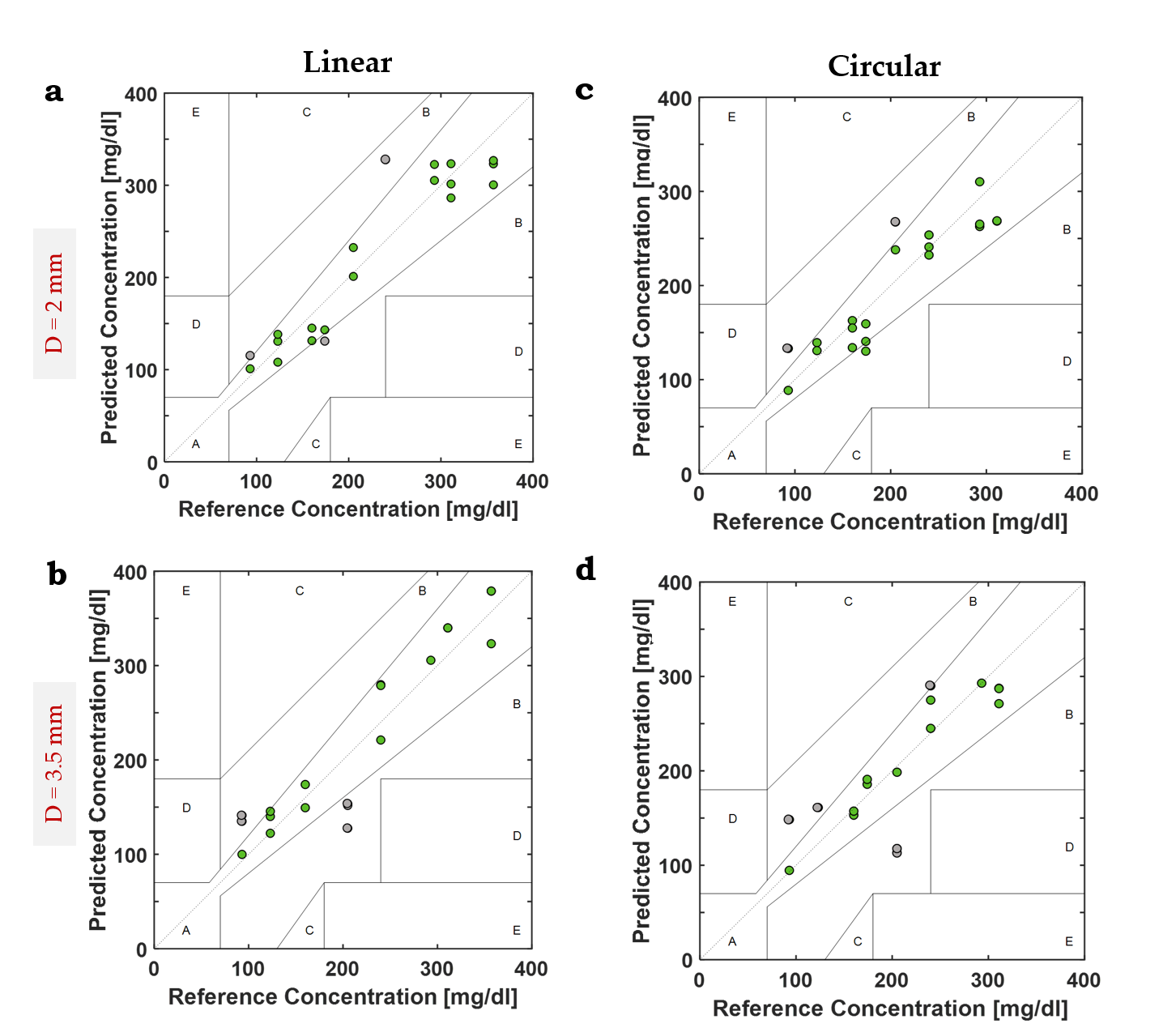}
    \caption{\textbf{CEGA for \textit{ex-vivo} experiments with P and R incidences:} \textbf{Top row}- shows the Clarke's Error Grid for the depth 2mm for P and R incidence respectively. \textbf{Bottom row}- shows the CEG while the chicken slice has a thickness of 3.5mm.}
    \label{fig:enter-label}
\end{figure}

\noindent
\newpage
\FloatBarrier

\bibliography{sample}